\documentclass[a4paper,onecolumn,11pt]{quantumarticle}
\pdfoutput=1
\usepackage[utf8]{inputenc}
\usepackage[english]{babel}
\usepackage[T1]{fontenc}
\usepackage{amsmath}
\usepackage{mathtools}
\usepackage{amssymb}
\usepackage{orcidlink}
\usepackage{hyperref}
\usepackage[numbers]{natbib}

\usepackage{tikz}
\usepackage{lipsum}

\newcommand{\la}{\langle}
\newcommand{\ra}{\rangle}
\newcommand{\ii}{\mathrm{i}}

\begin{document}

\title{Observing a $3T$ discrete time crystal on a trapped-ion qudit quantum processor}

\author{Gonzalo Camacho}
\email{gonzalo.camacho@dlr.de}
\affiliation{Department High-Performance Computing, Institute of Software Technology, German Aerospace Center (DLR), 51147 Cologne, Germany}
\orcid{0000-0001-6900-8850}

\author{Claire L. Edmunds}
\affiliation{Institut f{\"u}r Experimentalphysik, University of Innsbruck, Technikerstraße 25, 6020 Innsbruck, Austria}

\author{Michael Meth}
\affiliation{Institut f{\"u}r Experimentalphysik, University of Innsbruck, Technikerstraße 25, 6020 Innsbruck, Austria}

\author{Martin Ringbauer}
\affiliation{Institut f{\"u}r Experimentalphysik, University of Innsbruck, Technikerstraße 25, 6020 Innsbruck, Austria}
\orcid{0000-0001-5055-6240}

\author{Benedikt Fauseweh}
\email{benedikt.fauseweh@tu-dortmund.de}
\affiliation{Department of Physics, TU Dortmund University, Otto-Hahn-Str. 4, 44227, Dortmund, Germany}
\affiliation{Department High-Performance Computing, Institute of Software Technology, German Aerospace Center (DLR), 51147 Cologne, Germany}
\orcid{0000-0002-4861-7101}

\maketitle

\begin{abstract}
  Time crystals have been observed in various qubit-based quantum platforms. However, the realization of time-crystal behavior beyond period doubling has remained fairly unexplored, in part because established qubit architectures natively encode two-cycle dynamics. Qudits offer a natural route beyond this restriction. Here we propose a one-dimensional, disorder-free $S=1$ Floquet model with short-range interactions that realizes a discrete $3T$ time crystal and implement it on a trapped-ion qudit quantum processor. We observe period tripling dynamics in local observables and spin correlations, confirming the collective subharmonic response of the system in the experiment. The stabilization mechanism is analyzed by deriving the effective Floquet Hamiltonian and performing numerical simulations that demonstrate the existence of a prethermal phase over a wide range of parameters. We compute the phase diagram and verify the presence of multipartite entanglement through the Quantum Fisher Information, showing that this quantity gets enhanced at the crossover between ergodic and localized regimes in non-equilibrium.
\end{abstract}

\section{Introduction}


Periodically driven quantum systems open a vast landscape to explore the emergence of distinct non-equilibrium phases of matter~\cite{Basov2017,Moessner2017nat,Oka2019,Rudner2020}. In addition, they also pose some of the most fundamental and challenging questions in understanding emergent physical phenomena in many-body systems. In the absence of symmetries and in the presence of interactions, an isolated, periodically driven quantum system tends to absorb energy from the drive following the time evolution of the initial state. As a consequence, subregions of the system that can be locally probed are expected to evolve towards an effective infinite temperature state with trivial correlations~\cite{Polkovnikov2011,Alessio2014,Ponte2015anphys}. This idea is captured by the Eigenstate Thermalization Hypothesis~\cite{Deutsch1991,Srednicki1994} (ETH), which argues that generic quantum systems undergoing unitary evolution for sufficiently long times can eventually be well described by classical statistical ensembles~\cite{Rigol2008}. In such scenario, all knowledge about the initial conditions of the system is ultimately lost. Thus, systems satisfying the ETH are often termed ergodic, meaning that they uniformly explore all microscopic states in the course of time evolution. 

Remarkably, time crystals offer an exception to this generic ergodic fate by undergoing spontaneous symmetry breaking of time translational invariance~\cite{Wilczek2012,sacha_time_2018,khemani_brief_2019,else_discrete_2020,zaletel_colloquium_2023}, a phenomenon that can only occur under non-equilibrium conditions~\cite{watanabe_absence_2015}. In particular, discrete time crystals (DTCs) break the discrete periodicity $T$ of the external drive by undergoing collective subharmonic motion in their dynamics, with several spatio-temporal observables displaying $nT$-periodic motion with $n>1$. In the absence of dissipation, there are two well established mechanisms leading to time crystal stabilization: many-body localization (MBL)~\cite{Gornyi2005,Basko2006,Nandkishore2015} and prethermalization~\cite{Abanin2017,Mori2018,HO2023169297,Mallayya2019,Machado2020}. 

Under periodic driving, MBL prevents fast heating absorption due to the emergence of an extensive number of symmetries~\cite{Ponte2015}, with spontaneous symmetry breaking of a discrete time symmetry resulting in the emergence of a MBL discrete time crystal (MBL-DTC)~\cite{Khemani2016,Else2016,Ippoliti2021}. Experimental observations of the MBL-DTC phase~\cite{Zhang2017,Choi2017} have been replicated recently on superconducting qubits devices~\cite{Mi2022,Frey2022} and NV-center devices~\cite{randall_many-bodylocalized_2021}, further motivating the use of quantum hardware in the quest to identify genuine non-equilibrium quantum states~\cite{Fauseweh2024}. However, the experimental realization of MBL-DTCs on current quantum processors comes at the cost of including strong disorder sampling, which ultimately leads to averaging over a vast number of configurations, introducing a large overhead. 

Prethermalization prevents heating absorption in the regime of fast periodic drives, i.e. when the period $T$ of the drive satisfies $\lambda T\ll 1$ for some characteristic local energy scale of the system $\lambda$. In this regime the system can be described by an effective time-independent Hamiltonian governing the time evolution. Starting from an initial configuration with low effective energy will push heating effects to appear at exponentially long times, and the system resembles equilibrium at stroboscopic times. In addition, the effective Hamiltonian can contain emergent symmetries that are protected by the discrete time translational invariance of the drive. If the initial state breaks any emergent symmetry of the effective Hamiltonian, the state will cycle through the degenerate subspace of the broken symmetry~\cite{Else2017prx}, leading to a prethermal discrete time crystal. Most experimental observations of prethermal discrete time crystals refer to models containing long-range interactions, with implementations on qubit-based trapped-ion devices~\cite{Kyprianidis2021}, superconducting qubits~\cite{ying2022,solfanelli2024}, NV-center devices~\cite{beatrez_critical_2023} and NMR quantum emulator~\cite{Stasiuk2023}. However, previous theoretical work supports the existence of prethermal DTCs in one-dimensional models with short-range interactions~\cite{zeng2017,Luitz2020}. 

Another important aspect of DTCs is the $nT$ periodicity characterizing the dynamics. So far, experiments on DTC for $n=2$ have relied on long-range interacting systems~\cite{Zhang2017,Rovny2018}, short-ranged one-dimensional systems undergoing MBL~\cite{Mi2022,Frey2022} or alternative disordered setups~\cite{surace2019}. In order to realize $nT$-periodic DTCs with $n>2$, theoretical proposals relying on long-range interactions~\cite{liu2019,Machado2020,Pizzi2021,collura2022,munoz2022}, dissipative systems~\cite{Taheri2022} and mean-field approaches in ensembles of ultra-cold atoms~\cite{kuros_2020} have been developed. So far DTC behavior with $n>2$ periodicity has been experimentally reported in systems with long-range interactions~\cite{Choi2017} and ladder models~\cite{chen2026}. However, realizing DTCs beyond period-doubling on qubit-based architectures posses an obstacle given that qubits natively encode doubling period oscillations. Along these lines, qudit-based platforms are gaining considerable popularity for implementing DTC models beyond period doubling in a natural way~\cite{ma2025quditnativeframeworkdiscretetime}.

In this work, we investigate a one-dimensional, disorder-free $S=1$ Floquet model with nearest neighbour interactions whose unitary evolution operator induces a robust period-tripling response, even beyond perturbative regimes of the driving period. The existence of a collective tripling response in the model is verified by experiments carried out on a trapped-ion qudit quantum processor~\cite{Ringbauer2022} by investigating averaged magnetization and spin correlations.

We report that the model undergoes robust localized dynamics in Hilbert space for a wide range of parameters, a mechanism reminiscent of dynamical localization~\cite{Prosen1998,DALESSIO201319,luitz_absence_2017} and many-body scars~\cite{Bernien2017,Turner2018,Wen2019,Choi2019prl,Schechter2019,Chandran2023} based on the effective decoupling of Hilbert space sectors~\cite{Serbyn2021}. To further understand the observed behavior, we show that the choice of uniform and translational-invariant initial states prevents fast thermalization given that these are the low energy, three-fold degenerate symmetry broken states of the effective Floquet Hamiltonian. We focus on spatio-temporal correlations and bipartite entanglement dynamics, exploring the non-equilibrium phase diagram through numerical simulations that demonstrate the existence of a prethermal phase up to an exponentially long thermalization time scale. The non-equilibrium crossover between localized and delocalized dynamics is marked by an enhancement of multipartite entanglement.

\section{Preliminaries}

\subsection{Model and experimental realization}\label{subsec:model}

\begin{figure}[ht]
\centering
\includegraphics[width=\linewidth]{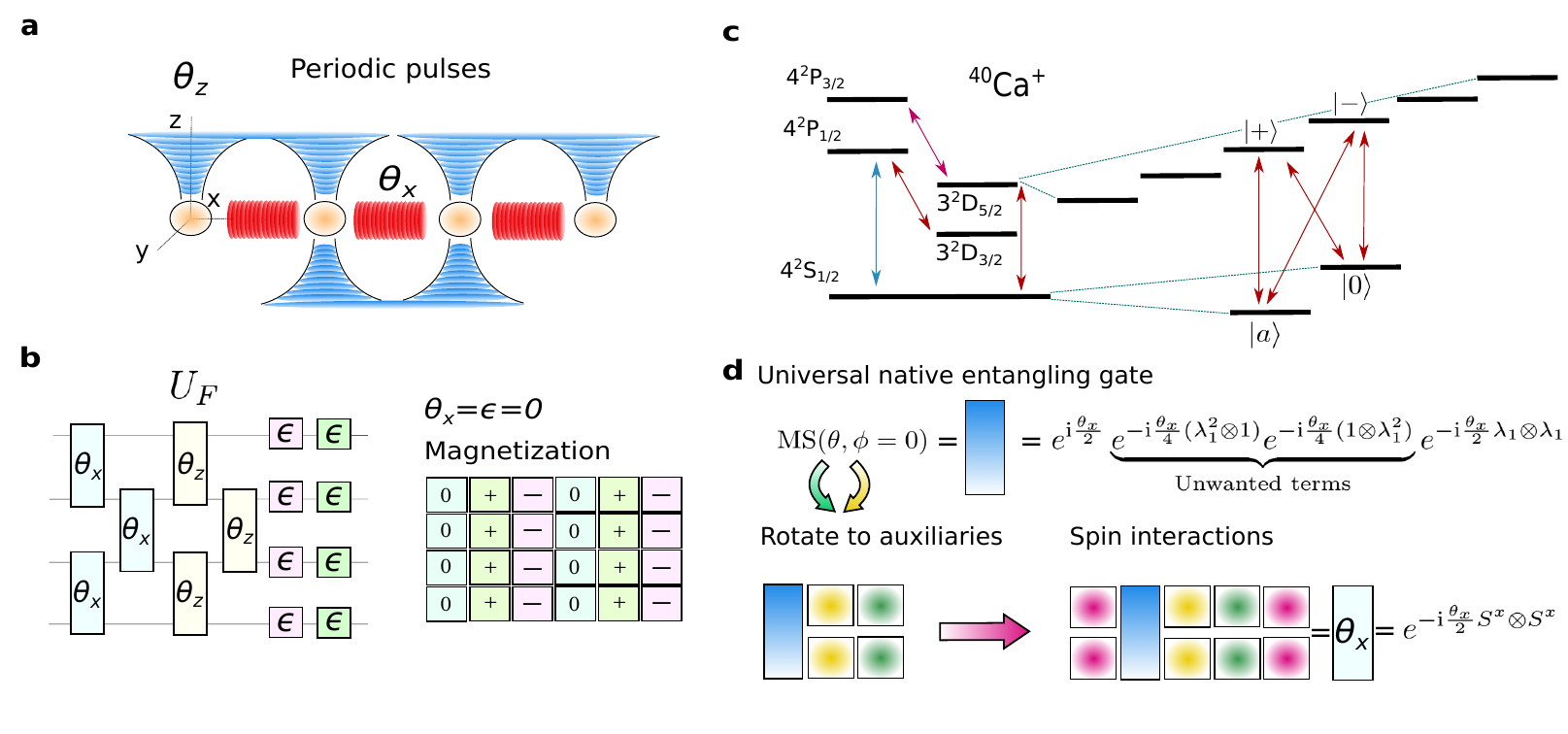}
\caption{{\bf{a}} Schematic representation of periodic pulses on a chain of trapped ions by alternating couplings $\theta_x,\theta_z$ in the $x$ and $z$ directions of the ions, respectively. {\bf{b}} The corresponding circuit realization for a digital quantum gate implementation of the Floquet unitary operator $U_F$ studied in this work. The kick parameter $\theta_x$ represents the strength of the entangling gates along the $x$ directions of spins, followed by a layer of entangling gates acting along the $z$ components of the spins depending on the parameter $\theta_z$. A set of local rotations represented by $P_\epsilon^{\mathbb{Z}_3}$ in Eq.~\eqref{eq:Floquet_un} follow the entangling gates. If both $\epsilon=\theta_x=0$, any initial product state in the $z$ basis of spins becomes an eigenstate of $U_F^3$, thus featuring a 3-periodic pattern in the magnetization; in the example, the initial state is $|\Psi_0\rangle=|0\rangle^{\otimes L}$ for a chain of $L=4$ spins. A finite value of $\theta_x$ will destabilize the perfectly oscillating pattern of the magnetization $\langle S^z\rangle$; employing entangling gates along the $z$ direction with strength $\theta_z$ right after the kicks along the $x$ direction contributes to restore the oscillations. {\bf{c}} Implementation of spin-spin interactions in the $^{40}\text{Ca}^+$ trapped-ion qudit quantum processor of Ref.~\cite{Ringbauer2022}, schematically showing the ion energy levels. The encoding of a single spin-1 into a four-dimensional space with basis states $|a\rangle,|+\rangle,|0\rangle,|-\rangle$ is sketched, along with the addressed transitions needed; note that our trapped-ion device can access more transitions between levels which are not represented here. {\bf{d}} The universal entangling gate $\text{MS}(\theta,\phi=0)$ contains unwanted local terms that can be isolated into the auxiliary level $|a\rangle$ of the encoding employing appropriate local rotations (Different rotation angles have been represented by different colors, see Appendix~\ref{sec:model_quditproc} for details). Isolation of unwanted local phases into the auxiliary levels allows to engineer spin-spin interacting gates for the Floquet operator $U_F$.}
\label{fig:fig1}
\end{figure}


We consider a one dimensional chain of $L$ spins with spin value $S=1$ interacting to nearest neighbors; unless otherwise stated, we employ open boundary conditions throughout this work, and use units of $\hslash=1$. As a local basis, we choose the eigenstates of the $S^z$ spin component, given by $\{|+\rangle,|0\rangle,|-\rangle\}$. The initial state is chosen to be $|\Psi_0\rangle=|0\rangle^{\otimes L}$. The state is subject to unitary evolution, which introduces interactions between neighbouring spins in the lattice by applying consecutive periodic pulses in the $x$ and $z$ directions, as sketched in Fig.~\ref{fig:fig1}{\bf{a}}. The periodic evolution of the state employs a ternary Floquet drive, represented by a unitary operator $U_F$ given by:
\begin{eqnarray}\label{eq:Floquet_un}
U_F&=&P_\epsilon^{\mathbb{Z}_3}e^{-\ii\frac{\theta_z}{2} \sum_{j=1}^{L-1} S^z_j S^z_{j+1} }e^{- \ii\frac{\theta_x}{2} \sum_{j=1}^{L-1} S^x_j S^x_{j+1} }=P_\epsilon^{\mathbb{Z}_3} U_z U_x,\nonumber\\
P_\epsilon^{\mathbb{Z}_3} &=&\prod_{j=1}^L e^{-\ii\frac{\pi-\epsilon}{2}\lambda_{6,j}}\prod_{j=1}^L e^{-\ii\frac{\pi-\epsilon}{2}\lambda_{1,j}},\nonumber\\
\lambda_{1}&=&\left(\begin{matrix}
0 & 1 &0\\
1 & 0 &0\\
0 & 0 &0
\end{matrix}\right),\hspace{5pt}\lambda_{6}=\left(\begin{matrix}
0 & 0 &0\\
0 & 0 &1\\
0 & 1 &0
\end{matrix}\right).
\end{eqnarray}
A digital circuit implementation of this Floquet unitary operator is represented in Fig.~\ref{fig:fig1}{\bf{b}}. The matrices $\lambda_{1,j},\lambda_{6,j}$ represent the first and sixth Gell-Mann matrices at site $j$ of the chain, respectively (following the notation of Ref.~\cite{Ringbauer2022}). 

The first two contributions of the drive, $U_x$ and $U_z$, consist of uniform spin-spin interactions between neighbouring sites, first with $U_x$ acting along the $x$ direction of the spins, and then with $U_z$ acting along the $z$ direction. The parameters $\theta_x,\theta_z$ represent the duration of the interaction pulses between the spins, which is scaled by the characteristic time scale of the uniform interactions. For a drive of period $T$, $\theta_x=J_xT_1$ and $\theta_z=J_z(T-T_1)$, with $J_x,J_z$ the spin interaction strength on each direction, and $0<T_1<T$. The $P_{\epsilon}^{\mathbb{Z}_3}$ term of the drive is considered to be instantaneous (a kick), and generates a set of local rotations featuring a discrete $\mathbb{Z}_3$ symmetry in the case $\epsilon=0$, with level transitions given by $|0\rangle\to |+\rangle,|+\rangle\to |-\rangle, |-\rangle\to |0\rangle$; for $\epsilon=\pi$, the $\mathbb{Z}_3$ symmetry in $P_{\epsilon=\pi}^{\mathbb{Z}_3}=\hat{1}$ is absent. In what follows, and unless otherwise stated, we will focus on the case $\epsilon=0$; the effect of finite $\epsilon$ values can be found in Appendix~\ref{app:eps_var}. If both $\epsilon=\theta_x=0$, any initial product state in the $\{|+\rangle,|0\rangle,|-\rangle\}$ basis is an eigenstate of $U_F^3$, and $\left|\langle \Psi_0|U_F^{3n}|\Psi_{0}\rangle\right|^2=1$ for any $n\in\mathbb{N}$, see Fig.~\ref{fig:fig1}{\bf{b}}.  Thus in this limit, and due to the choice of $|\Psi_0\rangle=|0\rangle^{\otimes L}$, subsequent applications of $U_F$ rotates the states within the set of uniform states $S_u=\{|+\rangle^{\otimes L},|0\rangle^{\otimes L},|-\rangle^{\otimes L}\}$. These states are approximate energy eigenstates in the high-frequency regime of the full model, remaining protected from small perturbations by the $\mathbb{Z}_3$ symmetry. The effective Floquet Hamiltonian $\bar{V}$ can be obtained in the interaction picture of the drive \cite{Else2017prx}. In the high frequency regime $U_F\approx P_0^{\mathbb{Z}_3}e^{-i\bar{V}T}$, we find,
\begin{eqnarray}\label{eq:Heff}
\bar{V}=\text{const.} + \frac{\theta_z}{T}\sum_{j} P_{j,j+1}+\bar{V}_x ,
\qquad P_{j,j+1}=\sum_{\sigma=+,0,-}|\sigma_j,\sigma_{j+1}\rangle\langle\sigma_j,\sigma_{j+1}|,
\end{eqnarray}
and $\bar{V}_x$ being a quasi-local perturbation proportional to $\theta_x$ hosting at most nearest-neighbour interactions. For details on the derivation of the effective Hamiltonian we refer to Appendix~\ref{app:effh}.

In the absence of $\theta_x$ perturbations, the states in $S_u$ are gapped (under periodic boundary conditions) by $\Delta_u= \frac{2\theta_z}{T}$ with respect to the rest of the (highly-degenerate) spectrum states, and are the only states of the spectrum being three-fold degenerate. Since $\bar{V}_x$ is quasi-local, this implies that low-order perturbation theory cannot in general couple the states in $S_u$, because a global flip of all spins is necessary. Due to the form of $\bar{V}_x$, an order-$n$ perturbation theory with $n\sim L$ is needed in order to couple the states in $S_u$, resulting in thermalization times scaling exponentially with $L$. Indeed, we have $\langle \Psi|\bar{V}_x|\Psi '\rangle=0$ for $|\Psi\rangle,|\Psi'\rangle\in S_u$. Thus, throughout the evolution by $U_F$ the system remains dynamically localized in the Hilbert-space region spanned by the states in $S_u$. As a consequence, due to $P_0^{\mathbb{Z}_3}$ any choice of $|\Psi_0\rangle\in S_u$ will feature prethermal time-crystalline behavior with period tripling in their dynamics. Indeed, since $\Delta_u\propto \theta_z$, larger values of $\theta_z$ will protect these uniform states from decaying throughout the evolution, resulting in a stronger localization effect in this region of the Hilbert space. 

We implement the model on a trapped-ion qudit quantum processor~\cite{Ringbauer2022}, where the spins of the chain are natively encoded as a \emph{qutrit}, i.e. three level quantum systems. We refer to Figs.~\ref{fig:fig1}{\bf{c}}, {\bf{d}}, and Appendix~\ref{sec:model_quditproc} for details on the encoding procedure.

\subsection{Implementation on the qudit quantum processor}\label{subsec:implementation}

The native entangling gate of the qudit processor in Ref.~\cite{Ringbauer2022} contains local terms that do not appear in the spin interactions present in the model (see~\cite{Ringbauer2022} and Appendix~\ref{sec:exp_setup} for details). To get rid of these unwanted phases and being able to implement the spin-spin interactions in Eq.~\eqref{eq:Floquet_un}, we encode each qutrit into an extended Hilbert space of dimension $4$, introducing an auxiliary degree of freedom that can be addressed by exploiting the allowed transitions between levels within a single ion, see Fig.~\ref{fig:fig1}{\bf{c}}.

 We define the encoding in a local basis $\{|a\rangle,|+\rangle,|0\rangle,|-\rangle\}$, where $|a\rangle$ represents the auxiliary level. The goal is to get rid of the local phases appearing in the native entangling gate~\cite{Ringbauer2022}:
\begin{eqnarray}\label{eq:ms_gate}
\text{MS}(\theta,\phi=0)=e^{\ii\frac{\theta}{2}}e^{-\ii\frac{\theta}{4}(\lambda_1^2\otimes 1)}e^{-\ii\frac{\theta}{4}(1\otimes \lambda_1^2)}e^{-\ii\frac{\theta}{2}(\lambda_1\otimes \lambda_1)},
\end{eqnarray}
with $\lambda_1$ being the first Gell-Mann matrix. The native single qutrit gates can be expressed in terms of the Gell-Mann matrices as:
\begin{eqnarray}\label{eq:rot_gates}
R_1(\theta,\phi) &=& e^{-\frac{\mathrm{i}\theta}{2}\left(\cos(\phi)\lambda_1 + \sin(\phi)\lambda_2\right)}=R_{+0}(\theta,\phi),\nonumber\\
R_2(\theta,\phi) &=& e^{-\frac{\mathrm{i}\theta}{2}\left(\cos(\phi)\lambda_4 + \sin(\phi)\lambda_5\right)}=R_{+-}(\theta,\phi),\nonumber\\
R_3(\theta,\phi) &=& e^{-\frac{\mathrm{i}\theta}{2}\left(\cos(\phi)\lambda_6 + \sin(\phi)\lambda_7\right)}=R_{0-}(\theta,\phi).
\end{eqnarray}
In the extended Hilbert space for the qutrit encoding, we can define local rotation gates $Z_{ij}(\theta)$ between levels $i,j$. For example, between the $|+\rangle,|0\rangle$ states:
\begin{eqnarray}
Z_{+0}(\theta)&=&R_1(\pi/2,-\pi/2)R_1(\theta,0)R_1(\pi/2,\pi/2)=\left(\begin{matrix}
1&0&0&0\\
0&e^{+\ii\frac{\theta}{2}}&0&0\\
0&0&e^{-\ii\frac{\theta}{2}}&0\\
0&0&0&1
\end{matrix}\right).
\end{eqnarray}
Thus, we can partially cancel the unwanted local terms as follows:
\begin{eqnarray}\label{eq:local_rot}
&\to&\left[\underbrace{Z_{a+}(-\theta)Z_{+0}(-\theta/2)}_{\text{qutrit} 1}\otimes \underbrace{Z_{a+}(-\theta)Z_{+0}(-\theta/2)}_{\text{qutrit}2}\right]\nonumber\\
&\times &\left[e^{\ii\frac{\theta}{2}}e^{-\ii\frac{\theta}{4}\left[(1\otimes \lambda_1^2)+(\lambda_1^2\otimes 1)\right]}e^{-\ii\frac{\theta}{2}(\lambda_1\otimes \lambda_1)}\right]\nonumber\\
&=&e^{\ii\frac{\theta}{2}}A_1 A_2 e^{-\ii\frac{\theta}{2}(\lambda_1\otimes \lambda_1)},
\end{eqnarray}
with $A_1,A_2$ being local operators acting on individual Hilbert spaces of qutrits 1 and 2, respectively, whose only non-trivial component belongs to the auxiliary level $|a\rangle$. Since the auxiliary level $|a\rangle$ does not couple under unitary dynamics to any other degrees of freedom, the unwanted phases will not affect the dynamics of the system. Alternatively, as the auxiliary rotations remove unwanted phases on qutrit sub-levels, they can also be replaced by phase tracking in software. In our work, we recorded the unwanted local phases on each qutrit level in software, and then updated the phases on all subsequent gates in the circuit to counteract the acquired phase. This approach is ideal to maximize fidelities across longer circuits.

This decomposition of the native gate allows to engineer spin-spin interactions in a simple way, by applying the appropriate local rotations to the exponential operator, transforming it to:
\begin{eqnarray}
    e^{\ii\frac{\theta}{2}}A_1 A_2 e^{-\ii\frac{\theta}{2}(\lambda_1\otimes \lambda_1)}\xrightarrow{\text{local rotations}}\begin{cases}
        e^{-\ii\frac{\theta_z}{2}S^z\otimes S^z},\\
        e^{-\ii\frac{\theta_x}{2}S^x\otimes S^x}.
    \end{cases}
\end{eqnarray}
To this end, the following relations are useful:
\begin{eqnarray}\label{eq:rot_lambdas}
R_2(\pi/2,\pi/2)\lambda_1 R_2^\dagger(\pi/2,\pi/2)&=&S_{x},\nonumber\\
R_1(-\pi/2,+\pi/2)\lambda_1 R_1(-\pi/2,+3\pi/2)&=&\lambda_3.
\end{eqnarray}
The exact form of the local rotations needed to transform the Gell-Mann matrices to spin-1 operators are described in detail in Appendix~\ref{sec:model_quditproc}.

\subsection{Quantum Fisher Information and numerical simulation details}

The Quantum Fisher Information (QFI) for a pure state $|\psi\rangle$ and observable $\mathcal{O}$ is given by:
\begin{eqnarray}\label{eq:qfi_pur}
    F_Q\left[|\psi\rangle,\mathcal{O}\right] &=&4\left(\langle \mathcal{O}^2\rangle - \langle \mathcal{O}\rangle^2\right),\nonumber\\
    \mathcal{O}&=&\sum_i \mathcal{O}_i.
\end{eqnarray}
As observables $\mathcal{O}$ we choose each of the total angular momentum components across all chain sites $\hat{J}_{l=x,y,z} =\sum_{i=1}^L S_i^l$, for which the corresponding $F_Q^l$ is calculated. The total QFI is defined as the sum over all components $F_Q =\sum_{l=x,y,z}F_Q^l$. For convenience, we work with the scaled quantity $f_Q=F_Q/8L$, since it allows for an easier visualization of the data compared against different multi-partite entanglement bounds (see Appendix~\ref{subsec:qfi_details} for details).

For finite chains up to $L=12$ spins, the time evolution of the state is carried out employing sparse matrix-vector multiplications. For the QFI in Fig.~\ref{fig:fig5}{\bf{c}}, we simulated the model employing the finite size TEBD algorithm~\cite{Vidal2003}, representing the operators as Matrix Product Operators (MPO). To address larger system sizes, we exploit the translational invariance of the model and employ the infinite version of the Time Evolving Block Decimation (iTEBD), keeping an error tolerance of $\varepsilon_{\text{TEBD}}=10^{-6}$, allowing for dynamical increment of the bond dimension every time the error threshold is exceeded, stopping the simulation whenever the bond dimension $\chi>600$.


\section{Results}

\subsection{Robust subharmonic oscillations}

\begin{figure}[ht]
\centering
\includegraphics[width=\linewidth]{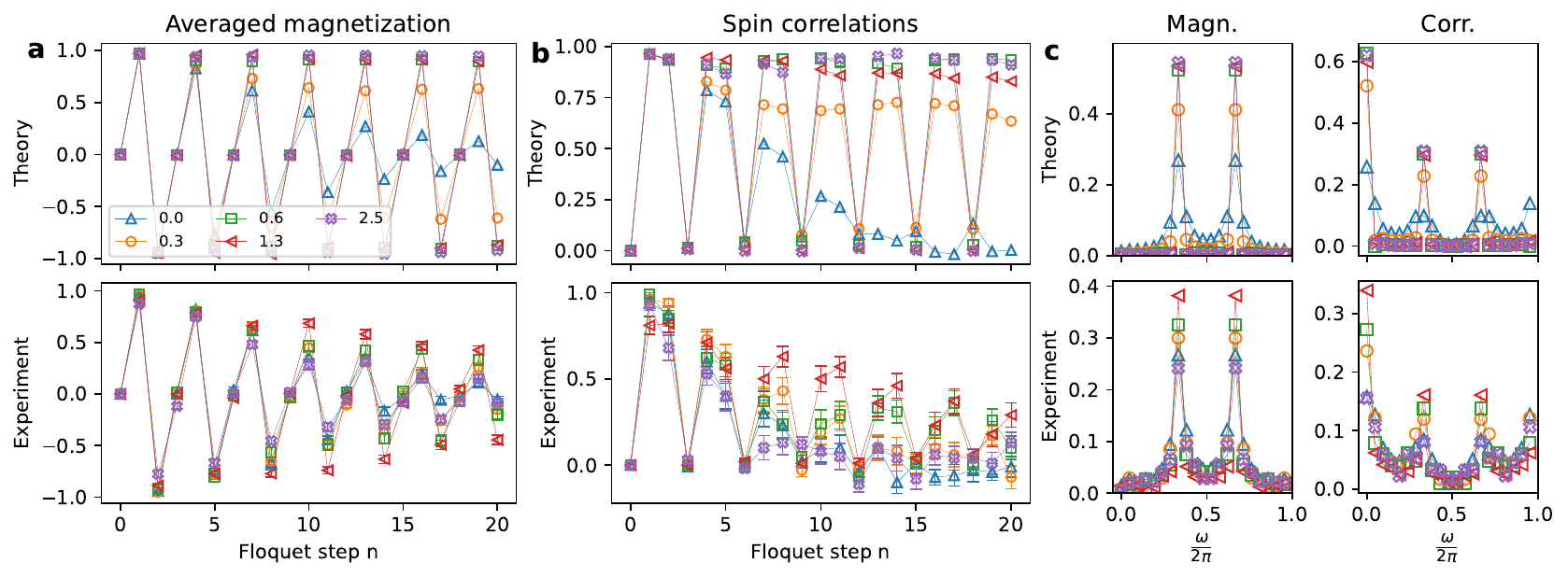}
\caption{ Comparison between theory and experiment results obtained in the trapped-ion qudit device of Ref.~\cite{Ringbauer2022} for a chain of $L=4$ spins, with $\theta_x=0.2$ and a total of $20$ Floquet steps, for different values of $\theta_z$ as indicated in the legend. Error bars corresponding to one standard deviation have been included in the experimental data, and are calculated using Monte Carlo resampling. {\bf{a}} The spatially averaged magnetization, showing the survival of oscillations with increasing value of $\theta_z$, signalling the appearance of dynamical localization. We observe that the experimental signal decays with increasing number of time steps due to noise, however we clearly see an enhancement with increasing $\theta_z$, with the exception of $\theta_z = 2.5$ which we attribute to coherent errors, see Appendix~\ref{sec:noisy_sims}. {\bf{b}} The spin correlations along the $z$ direction for two neighbouring spins in the chain get enhanced with increasing value of $\theta_z$. {\bf{c}} Discrete Fourier Transform representation to frequency domain for the magnetization and spin correlations from Figs. {\bf{a}} and {\bf{b}}. The appearance of peaked values in the Fourier spectrum at $\omega=\frac{2\pi}{3},\frac{4\pi}{3}$ indicates the emergence of subharmonic response to the external drive. }
\label{fig:fig2}
\end{figure}


\begin{figure}[ht]
\centering
\includegraphics[scale=0.8]{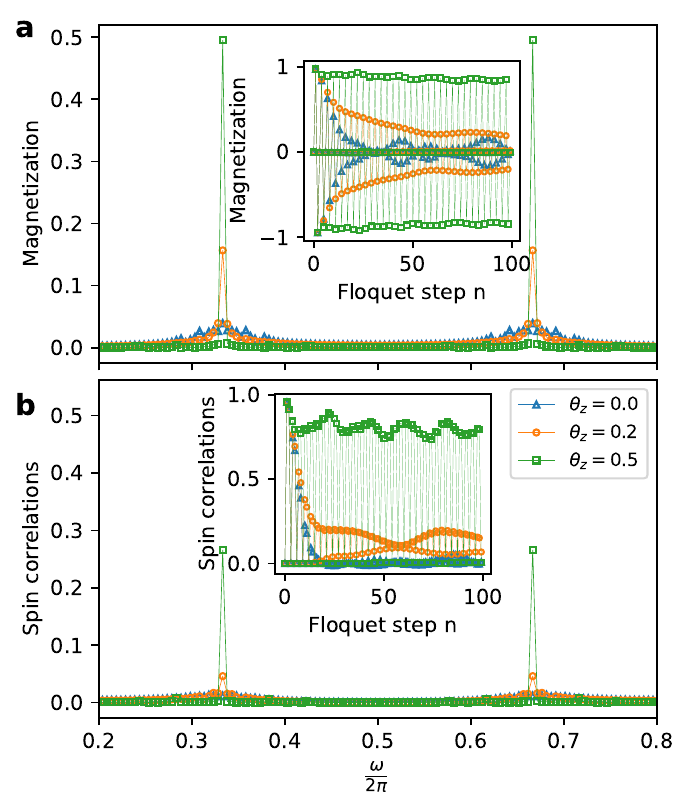}
\caption{Exact time evolution simulation for a chain of $L=12$ spins with $\epsilon=0$, $\theta_x=0.2$ and a total of $n=200$ Floquet steps. {\bf{a}} Fourier transform of the spatially averaged magnetization (see Appendix~\ref{sec:observables} for a precise definition of these observables); the insets represent the time evolution for each observable for the first $100$ steps of the evolution. {\bf{b}} Fourier transform of the spin-spin correlations in the $z$ direction between the middle chain spin and the right-most edge one. Both observables feature robust $3$-periodic oscillations with increasing $\theta_z$ values. }
\label{fig:fig3}
\end{figure}


We characterize the emergence of spatio-temporal order in the system dynamics by measuring the averaged magnetization across all spins of the chain and the spin correlations between spins located at different lattice sites. Figure~\ref{fig:fig2} shows the experimental results for a small chain of $L=4$ spins, compared to the theory prediction for a fixed parameter $\theta_x$ along the $x$ direction of spin interactions and for different values of the $\theta_z$ parameter. Our experimental results demonstrate that increasing values of $\theta_z$, i.e. stronger interaction between neighbouring spins along the $z$-direction, lead to the stabilization of the subharmonic response in the system, which otherwise quickly decays in the case of $\theta_z=0$.  
The fact that finite $\theta_z$ values following a kick with $\theta_x\neq 0$ restore the appearance of robust 3-periodic oscillations is a signature of dynamical localization in the system away from the trivial point $\theta_x=0$. The enhanced subharmonic response for finite $\theta_x$ is more clearly visualized in Fig.~\ref{fig:fig2}{\bf{c}}, where the Fourier representation of the associated time signal shows peaked values at frequencies $\frac{\omega}{2\pi}=1/3,2/3$. 

Note that the experiment does not achieve an exponentially long lifetime, but the lifetime is increasing for larger $\theta_z$, except for very large $\theta_z$, due to increasing gate errors. The time evolution is susceptible to noise, as was observed also in a recent realization of an $S=1/2$ MBL-DTC on a superconducting quantum processor \cite{Mi2022}. We estimated the gate errors and performed noisy simulations to confirm our noise model of the quantum processor, see Appendix~\ref{sec:noisy_sims}.

To verify the survival of these oscillations with increasing system size, and to rule out that the observed effect is solely due to revivals in the small system, we have complemented the results from Fig.~\ref{fig:fig2} with numerical simulation results for a chain of $L=12$ spins in Fig.~\ref{fig:fig3}. We conclude that both the magnetization and spin correlations experience a robust subharmonic response with increasing values of $\theta_z$ in larger system sizes, as evidenced by their representation in Fourier space. The persistence of subharmonic oscillations in the system is verified in the thermodynamic limit by exploiting the translational invariance of the system (see Appendix~\ref{sec:itebd}).

\subsection{Overlap and bipartite entanglement phase diagrams}


\begin{figure}[!ht]
\centering
\includegraphics[width=\linewidth]{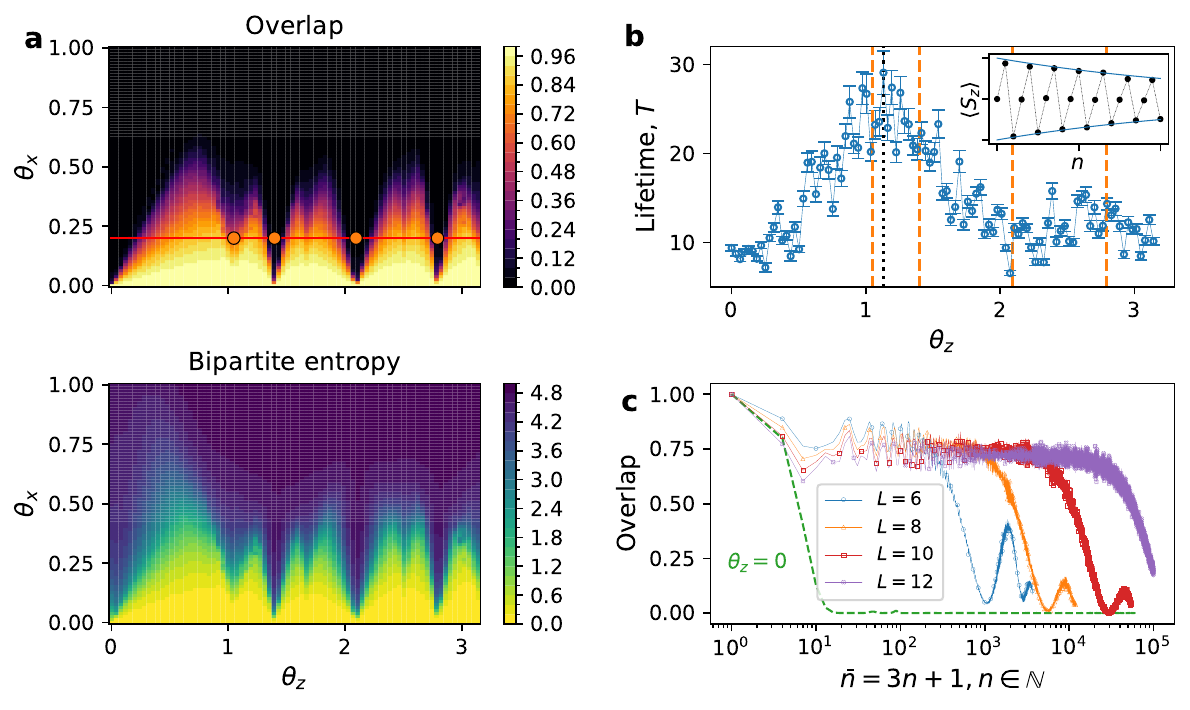}
\caption{{\bf{a}} Phase diagrams for the overlap and the bipartite entanglement entropy for a chain of $L=10$ spins, extracted from the exact time evolution of the state and averaged over a total of $500$ Floquet cycles, represented as a function of the pulse angles $\theta_x,\theta_z$. The overlap is averaged at time steps $\bar{n}=3n,n\in\mathbb{N}$. Note that times $\bar{n}$ correspond to values $\mathcal{F}=1$ for $\theta_x=0$. The dark region indicates ergodic behavior in the system, whereas bright values indicate a high probability of return to the initially prepared product state, signaling a localized regime. Values of $\theta_z=\frac{4\pi}{9},\frac{6\pi}{9},\frac{8\pi}{9}$ are identified for which the behavior is ergodic (dark regions) almost at any value of $\theta_x>0$. Both phase diagrams show a one-to-one correspondence with each other.
{\bf{b}} Contrast change in the experimental averaged magnetization data shown in Fig.~\ref{fig:fig2}{\bf{a}}. An exponential decay is fitted to the oscillation envelope to extract the lifetime $T$ in units of Floquet steps, $n$: $e^{-n / T}$. Error bars correspond to one standard deviation calculated using Monte Carlo resampling. (Inset) An example of an exponentially decaying envelope fitted to the average magnetization $\langle S_z\rangle$, shown for $\theta_z=1.1$, which corresponds to the maximum measured lifetime, $T=29(2)$ (black dashed vertical line in main figure). The location of the dips represented by the colored filled dots in Fig.~\ref{fig:fig4}{\bf{a}} correspond to the vertical dashed lines with matching color.
{\bf{c}} Large $n$ dependence of the overlap, with $\theta_x=0.2,\theta_z=0.5$ for different system sizes $L$; the dashed curve corresponds to $\theta_z=0,L=10$. To include the point $n=0$ in the logarithmic axis, data is represented against the shifted times $\tilde{n}=3n+1$, corresponding to the times for which $\mathcal{F}=1$ when $\theta_x=0$; note that the initial point in the logarithmic x-axis at $\bar{n}=1$ corresponds to the initial condition $n=0$. For data visualization purposes, the curves corresponding to $\theta_z=0.5$ have been represented only until times right after the overlap decays in value.
}
\label{fig:fig4}
\end{figure}

In order to explore the general behavior of the system for a wide range of variations in the $\theta_x,\theta_z$ parameters, we have represented the phase diagram for the averaged overlap of the time evolved state with the initial state and the half-chain entanglement entropy in Fig.~\ref{fig:fig4}{\bf{a}}, for a finite chain of size $L=10$. In both cases, we observe two well differentiated regions: A region where unitary dynamics presents localized behavior featuring high overlap with the initial state and small bipartite entropy, with the system retaining knowledge of the initial conditions in the state; a region showing diffusive dynamics corresponding to low overlap and high entropy, where knowledge of the initial configuration is erased. In the latter region, the dynamics of the system is ergodic. Interestingly, we observe values of $\theta_z$ for which the ergodic phase is present for almost any finite $\theta_x>0$; these values correspond to the dipped regions in Fig.~\ref{fig:fig4}{\bf{a}}. An analytic identification of two of these points is possible employing a simple perturbative expansion for the parameter $\delta_x=\frac{\theta_x}{2}$, as explained in Appendix~\ref{sec:app_perturbative}. 

Results obtained in the trapped-ion experiment for the contrast change in the site-averaged magnetization are represented in Fig.~\ref{fig:fig4}{\bf{b}}, with values of $\theta_z$ varying along the horizontal red line in Fig.~\ref{fig:fig4}{\bf{a}}. The observed lifetime of the average magnetization oscillations serves as a proxy to identify and contrast emergent ergodic and localized regions in the dynamics of the system, with peak lifetime values indicating high probability to return to the initial configuration at every 3-periodic application of the unitary $U_F$. In the experiment, two concurrent effects occur. First, the decoherence of the device leads to a temporal loss of signal. Second, thermalization in the ergodic regime overlaps with this decoherence, making it difficult to distinguish between the two phenomena. Despite these challenges, it is clearly evident that the decay exhibits a strong dependence on the localization parameter $\theta_z$. For values $\theta_z\sim 1.1$, the longest lifetimes are observed, in accordance with results from Fig.~\ref{fig:fig4}{\bf{a}}. Importantly, the experiment clearly reproduces the predicted dipped region at $\theta_z\sim 2.1$, observed in the phase diagrams in Fig.~\ref{fig:fig4}{\bf{a}}. This is followed by increased revivals of the initial states for $\theta_z>2.1$, thus verifying the existence of a dynamically localized regime. We note that this revival is observable despite increasing gate errors with $\theta_z$.

To further investigate the dependence with system size, we have represented the overlap and bipartite entanglement entropy in Figs.~\ref{fig:fig4}{\bf{c}} for different values of the chain size $L$ evolving the state up to very long time scales. We observe exponential scaling of the heating time with $L$ for the overlap with the initial state, which remains above $60\%$ for $L=12$ for a large number of cycles. This behavior is reminiscent of domain wall confinement effects occurring in some spin-$1/2$ kicked models~\cite{collura2022}, and whose stabilization mechanism is driven by longitudinal fields or long-range interactions. We note that for the model presented in this work, none of these features is present.

\subsection{Analytic estimation of the dips observed in the phase diagram}
We carried out a perturbative treatment in the parameter $\delta_x=\frac{\theta_x}{2}$ in order to estimate the thermalization scales by direct analytical calculation in a finite chain system. Discarding all contributions of order $\mathcal{O}(\delta_x^2)$ and higher, after three consecutive applications of $U_F$ over the initial state we obtain (up to normalization depending on $\delta_x$):
\begin{eqnarray}\label{eq:approx_state}
U_F^3|\Psi_0\ra\approx\left(|\Psi_0\rangle + e^{i\chi}|\Psi\ra+e^{i\chi}|\tilde{\Psi}\ra\right),
\end{eqnarray}
where $|\Psi\rangle,|\tilde{\Psi}\rangle$ are superpositions of states with zeros at every site except on two neighbouring sites, which contain a pair $+-$ or $-+$. Fixing the normalization constant for a total of $3n$ steps, this approach allows to obtain an approximate expression for the thermalization time scale, identified at the point where the overlap of the time evolved state respect to $|\Psi_0\rangle$ has value $1/2$. The total number of steps needed to reach that value in the overlap can be estimated to be the integer part of (see Appendix~\ref{sec:app_perturbative} for a detailed derivation):
\begin{eqnarray}\label{eq:nt_eq}
n_t&\sim&\frac{\log(1-\eta^2)-\log(1+\eta^2-2\eta\cos(\chi))}{\log\eta^2},\hspace{5pt}\eta=\frac{1}{\sqrt{1 + 2L\delta_x^2}},\hspace{5pt}\chi=\frac{9\theta_z}{2}.
\end{eqnarray}
Some maxima and minima for this function are identified at:
\begin{eqnarray}
\chi=m\pi\hspace{10pt}m\in\mathbb{Z}. 
\end{eqnarray}
In particular, for $m=0,2,4$ one obtains minima values of this function, corresponding to $\theta_z=0,\frac{4\pi}{9},\frac{8\pi}{9}$, respectively. The points $\theta_z=\frac{4\pi}{9},\frac{8\pi}{9}$ are in agreement with two of the dips observed in all phase diagrams at very small kick values of $\theta_x$. The dip at $\theta_z=\frac{6\pi}{9}$ corresponds to the case $m=3$. Given that the perturbative treatment identifies this point as a maxima rather than a minima, we argue that this point might originate from higher-order corrections.

\subsection{Multipartite entanglement non-equilibrium phase diagram}

\begin{figure}[!ht]
\centering
\includegraphics[width=\linewidth]{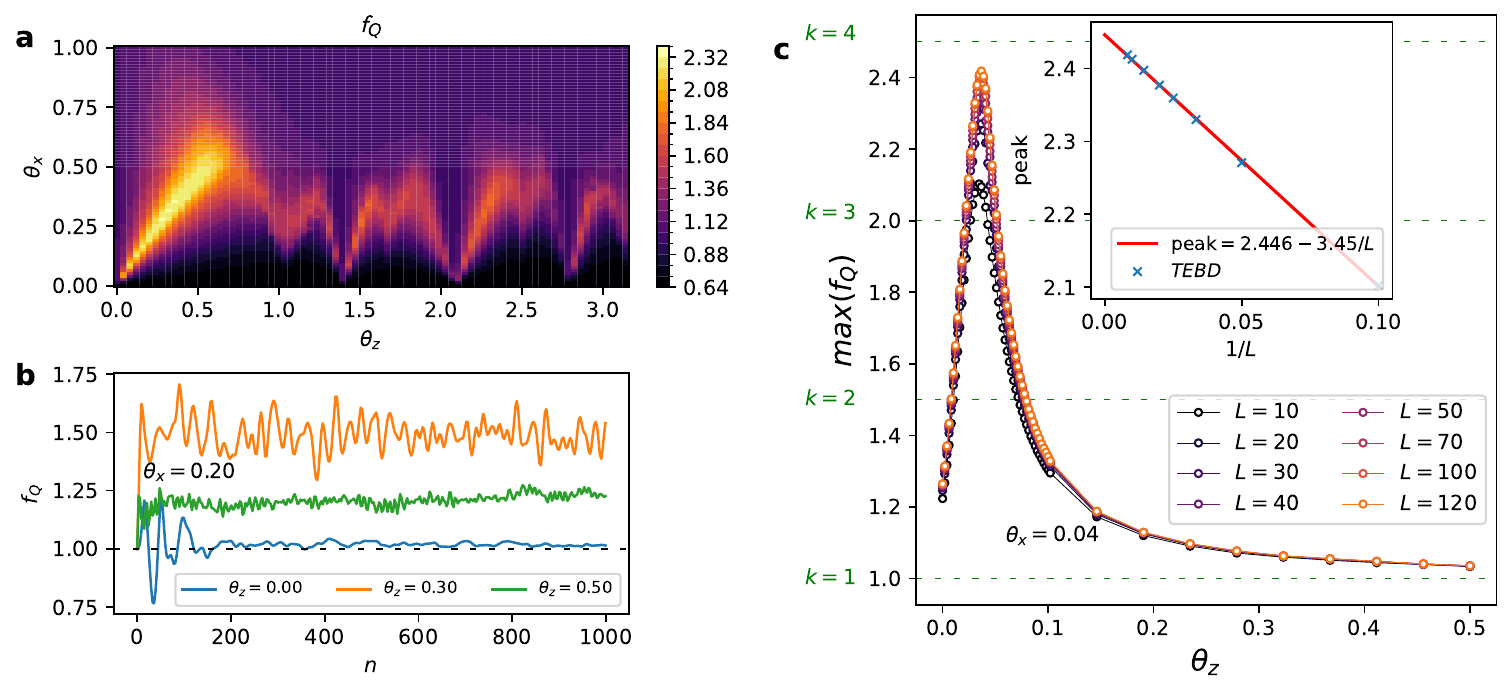}
\caption{{\bf{a}} Non-equilibrium phase diagram for the QFI obtain through exact time evolution on a chain of $L=10$ spin-1 spins, averaged over 500 Floquet cycles. The time-averaged QFI is enhanced at the boundaries in the crossover region separating the localized and ergodic regimes identified in Fig.~\ref{fig:fig4}{\bf{a}}. {\bf{b}} Time evolution of the QFI for fixed $\theta_x=0.2$ at times $\bar{n}=3n,n\in\mathbb{N}$. The dashed line indicates the upper bound in the QFI for separable states ($f_Q\leq 1$). 
{\bf{c}} (Main panel) The maximum value of the QFI for a total of $n=100$ Floquet cycles, with $\theta_x=0.04$ using the time evolving block decimation (TEBD) with tolerance $\epsilon_{\text{TEBD}}=10^{-6}$ for different system sizes $L$. The dashed horizontal lines represent different multipartite entanglement regions (with corresponding $k$-partite entanglement bounds represented by the dashed green lines), with $0.5\leq f_Q\leq 1$ for separable states; the peak of $f_Q$ is upper bounded by the 4-partite $(k=4)$ entanglement bound. {\bf{c}} (Inset) The scaling with system size $L$ of the peak in the QFI, showing the extrapolated value obtained in the $L\to \infty$ limit. The extrapolated result lies between the 3-partite and 4-partite upper bounds for entangled states (see the Appendix~\ref{subsec:qfi_details} for a definition of $f_Q$ and its entanglement bounds). }
\label{fig:fig5}
\end{figure}

As a genuine multipartite entanglement witness, the Quantum Fisher Information (QFI)~\cite{Pezze2018} has been shown to capture all relevant correlations leading to critical behavior for systems in thermal equilibrium, especially at zero temperature with ground states undergoing a quantum phase transition~\cite{Hauke2016}. The QFI has been previously explored in time crystal models undergoing dissipative dynamics, in particular the so called boundary time crystals~\cite{Lourenco2022,Montenegro2023}, as well as in disorder-free spin-$1/2$ models~\cite{yousefjani2025}. Here we employ this quantity to witness the presence of multipartite entanglement on the prethermal $S=1$ DTC model Eq.~\eqref{eq:Floquet_un} with no dissipation.

Figure~\ref{fig:fig5}{\bf{a}} shows the time-averaged scaled QFI $f_Q$ phase diagram for a finite chain of $L=10$ spins and a range of values for the kick parameters $\theta_x,\theta_z$. We show that the QFI clearly traces the transition from ergodic to nonergodic regions previously shown in Fig.~\ref{fig:fig4}, and that both phase diagrams are in accordance with each other. In particular, we observe a critical region where the QFI gets maximized near the 4-partite entanglement upper bound; this occurs when $\theta_x\sim\theta_z$. This behavior of multipartite entanglement, where the QFI increases considerably at the boundaries separating the localized and ergodic regions suggests an interpretation of the QFI as a generalized susceptibility measure to characterize the localization/delocalization crossover in periodically driven systems. To further complement these results, in Fig.~\ref{fig:fig5}{\bf{b}}, we have represented the dynamics of $f_Q$ for a fixed value of $\theta_x$ and three different $\theta_z$ values, monitoring the evolution of the QFI at times $\bar{n}=3n,n\in\mathbb{N}$, corresponding to the times where local rotations in the $\theta_x=0$ case bring the state back to its initial preparation, up to a global phase. A key result is the observation that $f_Q$ stabilizes to values corresponding to regions bounded by different multipartite entanglement sectors, and it does so within the time window plateau that identifies the duration of the prethermal phase in Fig.~\ref{fig:fig4}{\bf{c}}. For $\theta_z=0$, rapid thermalization occurs with the system showing multipartite entanglement. For finite $\theta_z$ values, dynamical localization locks the state to lie within different multipartite entanglement sectors. To further characterize the transition across the critical region witnessing maximal multipartite entanglement, in the main panel of Fig.~\ref{fig:fig5}{\bf{c}} we have represented the maximum value for the QFI for a small value of the kick parameter $\theta_x$, obtained from finite-size time-evolving block decimation (TEBD) simulations. We observe well-converged results at arbitrary $\theta_z$ values when approaching the thermodynamic limit by addressing larger system sizes $L$. The inset of Fig.~\ref{fig:fig5}{\bf{c}} shows the scaling of the peak with respect to system size, extrapolating its value in the thermodynamic limit $L\to\infty$. In this limit, we obtain a value for the peak of $\sim 2.446$, corresponding to a state between the 3-partite and 4-partite entanglement upper bounds. This results are again in accordance with those of the phase diagrams shown in Fig.~\ref{fig:fig4}{\bf{a}}.


\section{Outlook}\label{sec:outlook}
 
Our work exemplifies the potential of employing qudit-based quantum computers for studying discrete time crystals beyond period doubling, using local degrees of freedom that exceed those of conventional qubit-based processors. In contrast to many previously observed or theoretically proposed prethermal time crystals, our short-ranged interacting model applies beyond the fast periodic driving regime and does not require a dissipation mechanism. The dynamical localization is observed in the total magnetization as well as correlations and the crossover to the ergodic regime is marked by an enhancement of multipartite entanglement.

Since decoherence still posses a great obstacle in the realization of periodically driven systems in current digital quantum computers~\cite{Ippoliti2021,Mi2022,Frey2022}, correction methods based on quantum-classical feedback schemes~\cite{Camacho2024} could potentially help improve the observation of periodic dynamics for longer times.

The possibility of performing such experiments on quantum processors offers a unique opportunity to investigate quantum many-body dynamics undergoing anomalous thermalizing behavior~\cite{gyawali2024} beyond conventional qubit-based models. A fundamental understanding of the mechanism leading to dynamical localization is hence highly desirable.

We comment on generalizations of the work presented here, arguing that models with higher spin degrees of freedom could potentially share similar collective behavior in their dynamics whenever the unitary evolution operator involves a discrete $\mathbb{Z}_d$ symmetry, with $d=2S+1$ the local dimension associated to a system with spin value $S$. Another open question is to explore the stability of the prethermal phase against finite values of the local kick parameter $\epsilon$.

The role of the QFI as an experimentally accessible~\cite{Smith2016} witness of multipartite entanglement for systems in thermal equilibrium undergoing critical phase transitions is well known~\cite{Hauke2016}. Extending this notion away from thermal equilibrium offers great potential for unveiling universal non-equilibrium features of many-body quantum systems \cite{Baykusheva2023}. Our work shows that employing the QFI in closed DTC models undergoing prethermalization can further reveal fundamental links between the presence (absence) of multipartite entanglement and the emergence (disappearance) of a dynamical crossover between ergodic and localized phases, something that could have potential implications in the context of quantum metrology~\cite{Toth2012} beyond equilibrium.

\section*{Acknowledgements}
    This research was funded by the European Research Council (ERC, QUDITS, 101039522). Views and opinions expressed are however those of the author(s) only and do not necessarily reflect those of the European Union or the European Research Council Executive Agency. Neither the European Union nor the granting authority can be held responsible for them. We also acknowledge support by the Austrian Science Fund (FWF) through the SFB BeyondC (FWF Project No. F7109) and the EU-QUANTERA project T-NiSQ (I 6001-N), and by the IQI GmbH. The authors gratefully acknowledge the scientific support and HPC resources provided by the German Aerospace Center (DLR). The HPC system CARO is partially funded by ''Ministry of Science and Culture of Lower Saxony'' and ''Federal Ministry for Economic Affairs and Climate Action''. This work was
funded by the Deutsche Forschungsgemeinschaft (DFG, German Research Foundation)-Project No. FA
1884/5-1. Q-Neko project has received funding from the European Union’s Horizon Europe research
and innovation programme under Grant Agreement No. 101241875. This work was also performed
for Council for Science, Technology and Innovation (CSTI), Cross-ministerial Strategic Innovation
Promotion Program (SIP), “Promoting the application of advanced quantum technology platforms to
social issues”(Funding agency: QST)

\bibliographystyle{quantum}
\bibliography{refs}

\onecolumn
\appendix

\section{General relations}\label{sec:gen_relations}

The spin-1 operators in the local basis $\{|+\rangle,|0\rangle,|-\rangle\}$ are:
\begin{eqnarray}\label{eq:spin_defs}
S^z=\left(\begin{matrix}
1&0&0\\
0&0&0\\
0&0&-1
\end{matrix}\right),\hspace{10pt}S^x=\frac{1}{\sqrt{2}}\left(\begin{matrix}
0&1&0\\
1&0&1\\
0&1&0
\end{matrix}\right).
\end{eqnarray}
They satisfy the identities:
\begin{eqnarray}
(S^x_jS^x_{j+1})^3 = S^x_jS^x_{j+1},\hspace{10pt}(S^z_jS^z_{j+1})^3 = S^z_jS^z_{j+1}.
\end{eqnarray}
The exponential operators take the form:
\begin{eqnarray}
e^{-\ii\frac{\theta_\alpha}{2} S_j^\alpha S_{j+1}^\alpha}&=&1+\left(\cos\left(\frac{\theta_\alpha}{2}\right)-1\right)\left(S_j^\alpha S_{j+1}^\alpha\right)^2 -\ii \sin\left(\frac{\theta_\alpha}{2}\right)S_j^\alpha S_{j+1}^\alpha, \hspace{5pt}\alpha=x,z.
\end{eqnarray}

The Gell-Mann matrices $\lambda_1,\lambda_6$ are given by:
\begin{eqnarray}
    \lambda_1 =\left(\begin{matrix}
        0&&1&&0\\
        1&&0&&0\\
        0&&0&&0
    \end{matrix}\right),\hspace{5pt}\lambda_6 =\left(\begin{matrix}
        0&&0&&0\\
        0&&0&&1\\
        0&&1&&0
    \end{matrix}\right).
\end{eqnarray}
Note that $\left[\lambda_1,\lambda_6\right]\neq 0$. The individual rotations for $\epsilon=0$ are given by the matrices:
\begin{eqnarray}\label{eq:Pz3def}
P^{\mathbb{Z}_3}_{\epsilon=0}=e^{\ii\pi}\left(\begin{matrix}
0&\ii&0\\
0&0&\ii\\
1& 0&0
\end{matrix}\right).
\end{eqnarray}
The entangling gates of the model are invariant when changing $+\leftrightarrow -$ at $\epsilon=0$, i.e. under a  local unitary transformation; however, such transformation changes the local rotations term to:
\begin{eqnarray}\label{eq:Pz3def_rot}
\bar{P}^{\mathbb{Z}_3}_{\epsilon=0}=e^{\ii\pi}\left(\begin{matrix}
0&0&1\\
\ii&0&0\\
0& \ii&0
\end{matrix}\right).
\end{eqnarray}

\section{Observables}\label{sec:observables}
Under unitary dynamics, the state of the system remains pure at any arbitrary time step $n$; we represent it by $|\Psi_n\rangle$ for notational convenience. The main observable quantities addressed in this work are:\begin{itemize}
\item The spatially averaged spin magnetization at time steps $n$, defined by
\begin{eqnarray}
\langle \Psi_n| S^z|\Psi_n\rangle \coloneq \langle S^z(n)\rangle=\frac{1}{L}\sum_{j=1}^L \langle \Psi_n| S_j^z|\Psi_n\rangle,
\end{eqnarray}
with index $j$ labelling spatial indices on the lattice.
\item The spin-spin correlations between the spin located in the middle of the chain and the one located at the right-most edge $\langle \Psi_n|S^z_{L/2}S^z_{L}|\Psi_n\rangle$.
\item The overlap respect to the initial state, representing the return probability, is given at any time step by $\mathcal{F}(n)=|\langle\Psi_0|\Psi_n\rangle|^2$.
\item The mid-chain von Neumann entropy $S_{L/2}(n)$ for a bipartition $\mathrm{A},\mathrm{B}$ in the middle of the chain, defined as:
\begin{eqnarray}
    S_{L/2}(n)=- \text{tr}\left(\rho_n^\mathrm{A}\log \rho_n^\mathrm{A}\right), \hspace{10pt} \rho_n^\mathrm{A}=\text{tr}_\mathrm{B} \rho_n,
\end{eqnarray}
with $\rho_n=|\Psi_n\rangle\langle\Psi_n|.$

\item For a pure state $|\psi\rangle$, the QFI respect to an operator $\mathcal{O}$ is given by the variance:
\begin{eqnarray}
    F_Q\left[|\psi\rangle,\mathcal{O}\right] &=&4\left(\langle \mathcal{O}^2\rangle - \langle \mathcal{O}\rangle^2\right).
\end{eqnarray}
In the main text, we always employ $\mathcal{O}=\sum_i \mathcal{O}_i$ as generators composed by a sum of local operators. The generators are chosen to be those of angular momentum along $x,y,z$ directions of the spin, for which $\mathcal{O}^{\alpha=x,y,z}=\sum_i \mathcal{O}_i^\alpha.$

\end{itemize}

For the frequency space representation of the discrete time signals consisting of $N$ total time steps, we define the discrete frequencies $\omega_k,k\in\{0,1,...,N-1\}$ and use the discrete Fourier transform; for the spatially averaged magnetization:
\begin{eqnarray}
    \langle S^z(\omega_k)\rangle\coloneq \langle S^z(\omega)\rangle=\sum_{n=0}^{N-1}\langle S^z(n)\rangle e^{-\ii \frac{2\pi nk}{N}}.
\end{eqnarray}

\section{Noisy simulations}\label{sec:noisy_sims}

\begin{figure}[!ht]
    \centering
    \includegraphics[width=\linewidth]{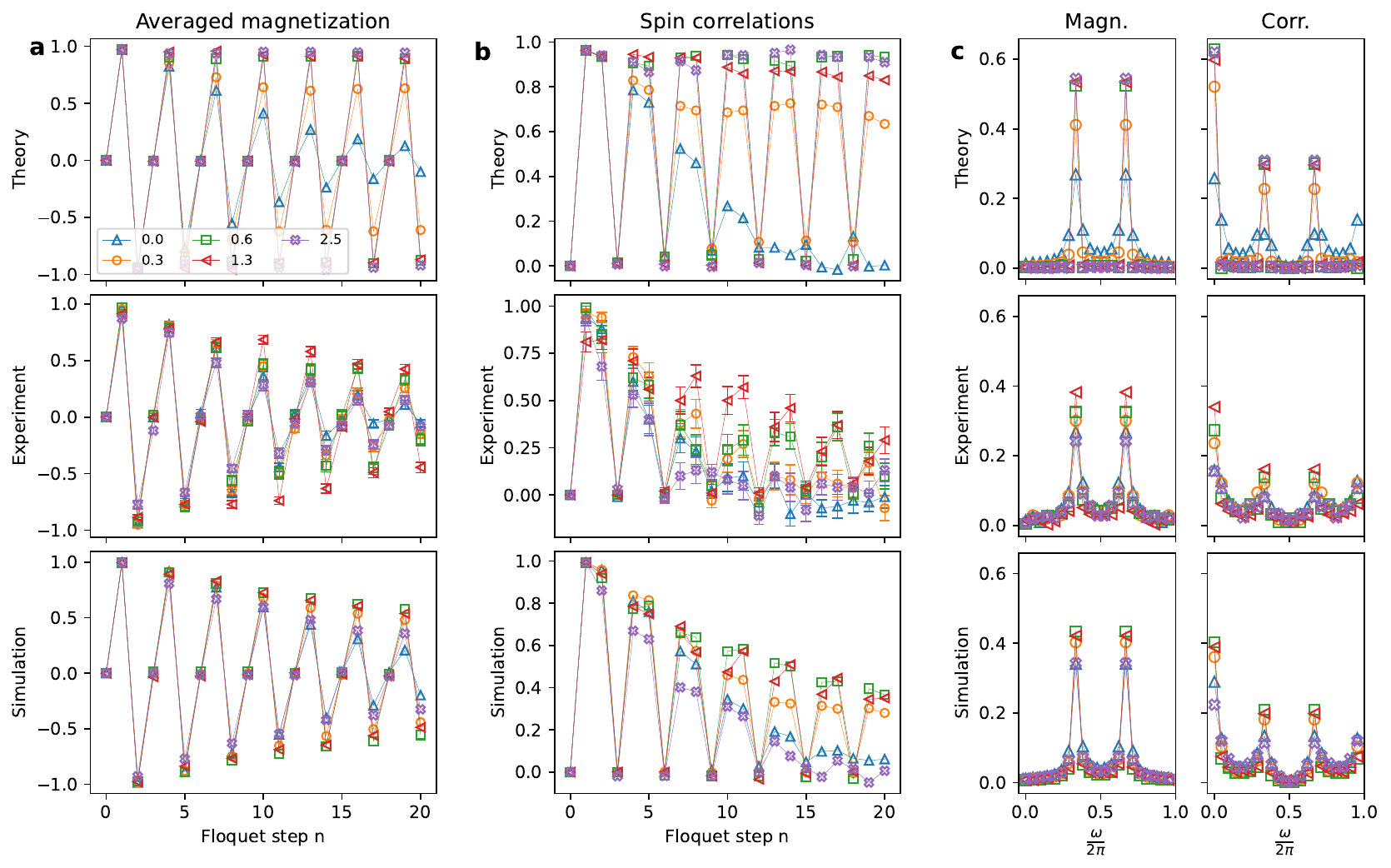}
    \caption{Same as Fig.2 in the main text, including the noisy simulation data in {\bf{a}} and {\bf{b}}. In the presence of noise, the enhancement of the period tripling both in the magnetization and the spin correlations is visible in the simulated data set with the exception of $\theta_z=2.5$, in agreement with the observed behavior in the experiment.  }    
    \label{fig:fig_s0}
\end{figure}

In this section, we include results on the noisy simulations carried out to verify the noise model of the qudit processor. 

The noisy simulations were performed by calculating the full state evolution with probabilistic Pauli bit and/or phase flip errors to capture depolarizing and dephasing error channels. The simulation was repeated 50 times with both depolarizing and dephasing error probability $2\theta\times0.03$ $(\theta = \theta_x, \theta_z)$, and 300ms $T_2$ time. Depolarizing errors were added probabilistically on the addressed two-level transition following an entangling gate (either the $S^x\otimes S^x$ or $S^z\otimes S^z$), and were randomly chosen to manifest as a x-, y-, or z-axis $\pi$-rotation. Additionally, entangling gates induce dephasing errors on the other qudit levels, which are simulated using a z-axis $\pi$-rotation (phase flip error) following an entangling gate. Finally, dephasing with $T_2=300$\,ms was modelled using phase flip errors on both the $\vert+\ra\leftrightarrow\vert0\ra$ and $\vert0\ra \leftrightarrow \vert-\ra$ two-level transitions. The errors occur probabilistically after the three Trotter step operators based on their duration: $P_\epsilon^{\mathbb{Z}_3}$ ($120\,\mu$s), $S^x\otimes S^x$ ($800\,\mu$s), and $S^z\otimes S^z$ ($560\,\mu$s).

The results are shown in Fig.~\ref{fig:fig_s0}, where the ideal case scenario (theory), the experiment and the noisy simulator are compared to each other.

\section{Persisting oscillations in the thermodynamic limit}\label{sec:itebd}

\begin{figure}[!ht]
    \centering
    \includegraphics[scale=0.7]{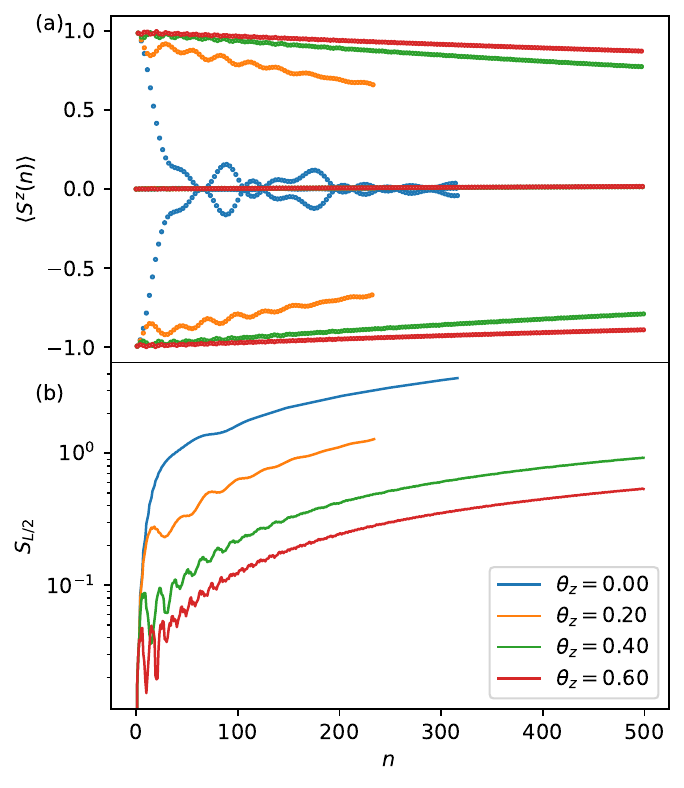}
    \caption{iTEBD results for $\theta_x=0.1$. The color scheme of the legend applies to both plots. (a) The averaged magnetization of the chain; lines connecting data points have been removed to allow for a better visualization. The system magnetization features robust 3-periodic oscillations for increasing $\theta_z$ values. (b) Growth of the bipartite entanglement entropy $S_{L/2}$ for different $\theta_z$. The $y$-axis has been represented in logarithmic scale. Bigger $\theta_z$ values lead to a slower growth of the bipartite entanglement in the thermodynamic limit.  }    
    \label{fig:fig_s1}
\end{figure}

\begin{figure}[t!]
    \centering
    \includegraphics[scale=0.7]{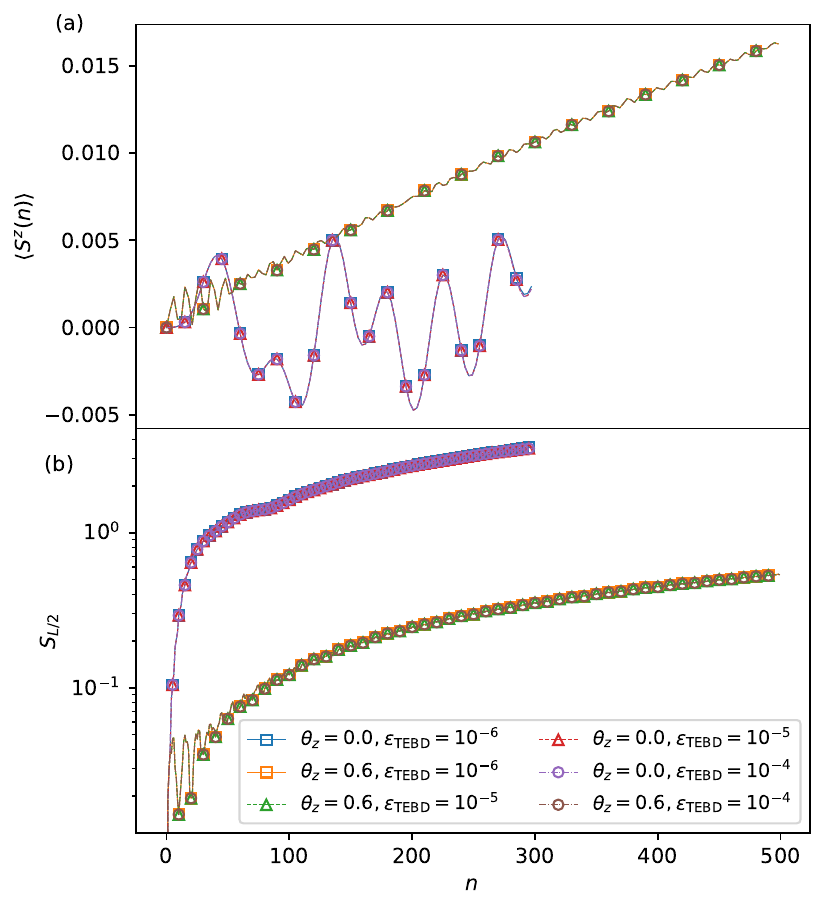}
    \caption{iTEBD calculations for different values of the tolerance parameter $\epsilon_{\text{TEBD}}$ and $\theta_x=0.1$. In (a) spatially averaged magnetization at times $3n$ starting from the $|0\rangle^{\otimes L}$ state. Notice the steady increment away from zero values for $\theta_z=0.6$. (b) Bipartite entanglement entropy $S_{L/2}$ at different time steps. Only few markers of data points have been included in the plot for data visualization purposes.  }    
    \label{fig:fig_s2}
\end{figure}

To verify the survival of the observed oscillations when approaching the thermodynamic limit, we have represented results obtained by the iTEBD algorithm employing a truncation error $\epsilon_{\text{TEBD}}=10^{-6}$ in Fig.~\ref{fig:fig_s1}, using different values of the $\theta_z$ parameter while keeping $\theta_x$ fixed. Increasing $\theta_z$ drives the state to oscillate periodically with period 3 times that of the application of the Floquet unitary $U_F$, prolonging the survival of oscillations. We stress that even when $\theta_x$ is finite, increasing $\theta_z$ leads to an overall behavior closer to the integrable point of the model. In Fig.~\ref{fig:fig_s1}(b), we show results for the time evolution of the mid-chain bipartite entanglement entropy $S_{L/2}$. For increasing values of $\theta_z$, there is an overall decrease in the bipartite entanglement growth, confirming the existence of dynamical localization for a wide range of Floquet cycles.

To check convergence in the observables calculated by iTEBD, in Fig.~\ref{fig:fig_s2} we have represented different iTEBD runs for different values of $\epsilon_{\text{TEBD}}$ at $\theta_z=0,0.6$ and $\theta_x=0.1$. A smaller value of $\epsilon_{\text{TEBD}}$ translates into a faster dynamical growth of the required MPS bond dimension $\chi$ in order to represent the state within the desired error tolerance.

\section{Initial state variation and deviations from perfect pulses}\label{app:eps_var}

One important feature characteristic of prethermal discrete time crystals is the dependence on different initial states. In general, only low-energy, highly ordered initial states of the effective Floquet Hamiltonian undergo time crystal dynamics, whereas random initializations will decay quickly~\cite{khemani_brief_2019,Ippoliti2021}.  

\begin{figure}[!ht]
    \centering
    \includegraphics[scale=0.85]{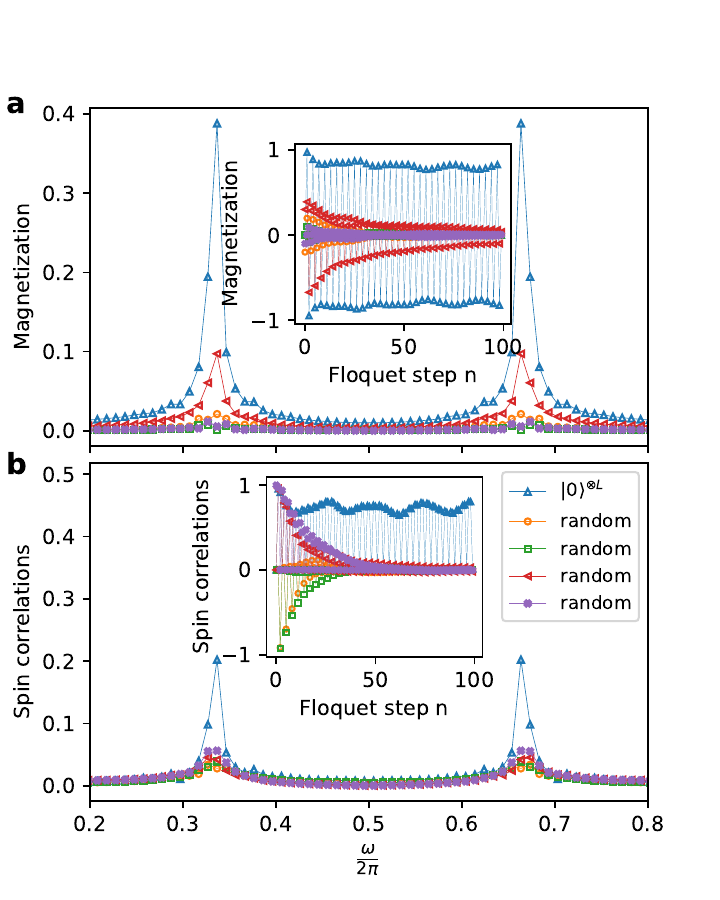}
    \caption{Initial state variation for a chain with $L=10$ spins, $\epsilon=0$, $\theta_x=0.2$, $\theta_z=0.4$, showing (a) the magnetization and (b) spin correlations, for different initial states specified in the legend. The fully polarized state undergoes robust tripling oscillations on both quantities (insets), featuring pronounced peaks in Fourier spectrum at values $\frac{\omega}{2\pi}=\frac{1}{3},\frac{2}{3}$, whereas initial random configurations in the $\{+,0,-\}$ basis rapidly decay with no structure in Fourier space.}    
    \label{fig:fig_s3}
\end{figure}

Here we compare the time evolution of the fully polarized initial state $|0\rangle^{\otimes L}$ respect to random initial product states $|s_j\rangle^{\otimes j},s_j\in\{+,0,-\}$. The results are shown in Fig.~\ref{fig:fig_s3}, evidencing that random initial configurations do not display time crystalline behavior, contrary to the fully polarized state. As discussed in the main text, fully polarized states represent low-energy states of the effective Floquet Hamiltonian.

\begin{figure}[!ht]
    \centering
    \includegraphics[scale=0.8]{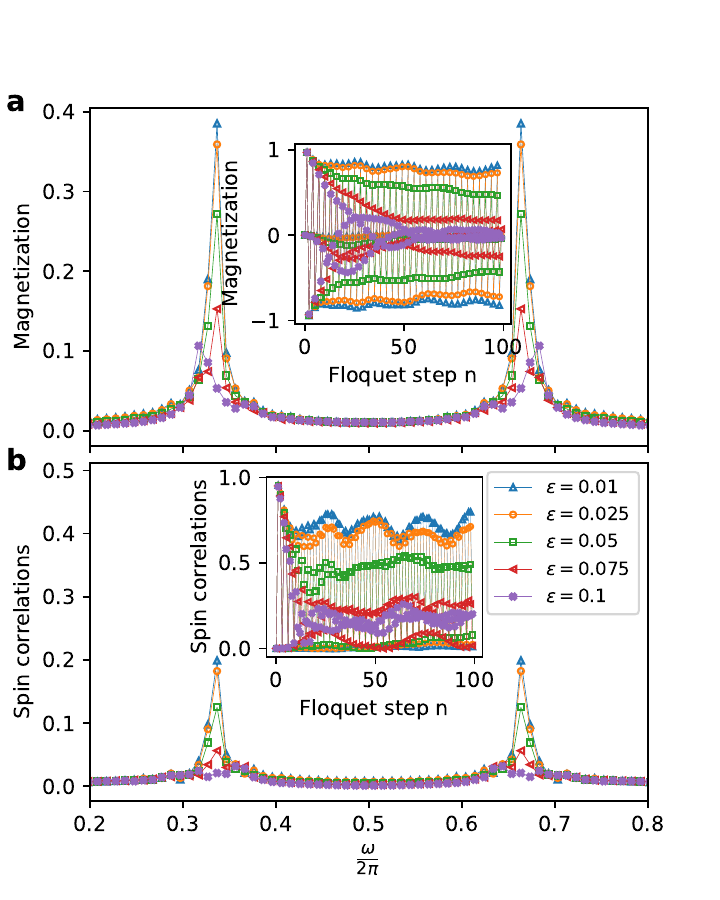}
    \caption{Effect of finite $\epsilon$ values for a chain with $L=10$ spins $\theta_x=0.2$, $\theta_z=0.4$, showing (a) the magnetization and (b) spin correlations starting from the fully polarized state $|0\rangle^{\otimes L}$. For small $\epsilon$ regimes, tripling oscillations remain robust signalling DTC-like behavior in this regime.}    
    \label{fig:fig_s4}
\end{figure}

Another aspect to consider for the model introduced in the main text is the effect of finite values $\epsilon>0$ deviating the application of single-qudit rotations from a perfect pulse. The stability of the DTC phase against these coherent deviations is of interest to show the robustness of the DTC even in the presence of coherent errors in the device.

We investigate the variation of the $\epsilon$ parameter away from the perfect pulse point $\epsilon=0$. The results are shown in Fig.~\ref{fig:fig_s4} for the fully polarized initial state $|0\rangle^{\otimes L}$. We observe that robust oscillations are preserved for small values of $\epsilon$, while gradually increasing $\epsilon$ drives the system from the DTC phase to a rather ergodic phase showing no oscillations. This demonstrates the survival of the time-crystalline phase even in the presence of small coherent errors deviating from perfect pulses.

\section{Details on perturbative dynamics}\label{sec:app_perturbative}

In this section, we provide a perturbative treatment to approximately obtain the time evolved state. To simplify the approach, we consider a chain of $L$ spins with PBC (note the choice of boundary conditions is irrelevant in the thermodynamic limit). In what follows we assume $\epsilon=0$ and $\theta_x\sim 0$, and change $+\leftrightarrow -$ under a local rotation of the spins, which leaves the entangling gates unchanged and local rotations are now given by Eq.~\eqref{eq:Pz3def_rot}.

In this regime, we treat the term including the kick $\theta_x$ in a perturbative way, discarding contributions of order $\theta_x^2$ or higher at any subsequent time steps. Note that this introduces an error after each application of the unitary. Starting from the initial state $|\Psi_0\rangle=|0\rangle^{\otimes L}$ we can write:
\begin{eqnarray}\label{eq:expH1psi0}
e^{-\ii\delta_x\sum_j S_{j}^xS_{j+1}^x}|\Psi_0\rangle &\sim& |0\rangle^{\otimes L}\nonumber\\
&-&\mathrm{i}\delta_x\sum_j |0...+-...0\rangle + |0...-+...0\rangle\nonumber \\
&+&\mathcal{O}(\delta_1^2),
\end{eqnarray}
up to a normalization constant, which is determined to be (under PBC):
\begin{eqnarray}\label{eq:N1_norm}
\mathcal{N}_1=\frac{1}{\sqrt{1+2L\delta_x^2}}, \hspace{10pt}\delta_x=\frac{\theta_x}{2}.
\end{eqnarray}
We write the approximated state in Eq.~\eqref{eq:expH1psi0} as:
\begin{eqnarray}
|\Psi_x\rangle &=&\mathcal{N}_1\left(|\Psi_0\rangle +|\Psi\rangle+|\bar{\Psi}\rangle\right),
\end{eqnarray}
where $|\Psi\rangle,|\bar{\Psi}\rangle$ are mutually orthogonal respect to each other and respect to the initial state. Application of the drive with $\theta_z$ introduces a phase (in what follows we use $\delta_z=\frac{\theta_z}{2}$), and a perfect pulse with $P^{\mathbb{Z}_3}_{\epsilon=0}$ is defined by Eq.~\eqref{eq:Pz3def}. After a single application of $U_F$, we find:
\begin{eqnarray}
U_F|\Psi_0\rangle &\propto& |-\rangle^{\otimes L}\nonumber\\
&-&\ii\delta_x e^{+\ii\delta_z}\sum_j|-...0_j,+_{j+1}-...\rangle\nonumber\\
&-&\ii\delta_x e^{+\ii\delta_z}\sum_j|-...+_j,0_{j+1}-...\rangle
\end{eqnarray}
We apply $U_F$ a second time, noting that now the application of the drive with $\theta_x$ leaves the state invariant only keeping terms of $\mathcal{O}(\delta_x)$. Application of the $\theta_z$ term and the local rotations gives:
\begin{eqnarray}
U_F^2|\Psi_0\rangle &\propto&
e^{-\ii L\delta_z}|+\rangle^{\otimes L}\nonumber\\
&-&\ii\delta_x e^{-i(L-5)\delta_z}\sum_j |+...-_j,0_{j+1}+...\rangle\nonumber\\
&-&\ii\delta_x e^{-\ii(L-5)\delta_z}\sum_j|+...0_j,-_{j+1}+...\rangle,
\end{eqnarray}
up to a normalization constant. The last application of $\theta_x$ again does not change the state in $\mathcal{O}(\delta_x)$. Therefore:
\begin{eqnarray}
U_F^3|\Psi_0\rangle &\propto& e^{-\ii 2L\delta_z}|0\ra^{\otimes L}\nonumber\\
&-&\ii\delta_xe^{-\ii(2L-9)\delta_z}\sum_j |0...+_j,-_{j+1}0...\rangle\nonumber\\
&-&\ii \delta_xe^{-\ii (2L-9)\delta_z}\sum_j|0...-_j,+_{j+1}0...\rangle.
\end{eqnarray}
We note that $e^{-\ii 2L\delta_z}$ is a global phase and thus can be dropped in what follows. After three Floquet cycles ($\hat{T}=U_F^3$), we have a state very similar to that of Eq.~\eqref{eq:expH1psi0}. Just the factor $\delta_x$ acquires a phase:
\begin{eqnarray}
\delta_x\to \delta_x e^{i \chi},\hspace{5pt}\chi= \frac{9\theta_z}{2}.
\end{eqnarray}
Calling $\hat{T}=U_F^3$, we can write the approximated state as:
\begin{eqnarray}
\hat{T}|\Psi_0\ra\approx \mathcal{N}_1\left(|\Psi_0\rangle + e^{i\chi}|\Psi\ra+e^{i\chi}|\tilde{\Psi}\ra\right).
\end{eqnarray}
To simplify notation, we call $|\delta\ra=|\Psi\ra+|\tilde{\Psi}\ra$. The state after $n$ applications of $\hat{T}$ in $\mathcal{O}(\delta_x)$ is, up to a normalization constant:
\begin{eqnarray}
\hat{T}^n|\Psi_0\rangle = \mathcal{N}_1^n|\Psi_0\ra + \underbrace{\left(e^{i(n+1)\chi}\sum_{k=1}^n \mathcal{N}_1^k e^{-ik\chi}\right)}_{=F_n}|\delta\ra.
\end{eqnarray}
The second coefficient is:
\begin{eqnarray}\label{eq:Fn_norm}
F_n &=&\frac{e^{i\chi}\mathcal{N}_1\left(e^{in\chi}-\mathcal{N}_1^n\right)}{e^{i\chi}-\mathcal{N}_1}\nonumber\\
|F_n|^2&=&\mathcal{N}_1^2\left(\frac{1+\mathcal{N}_1^{2n}-2\mathcal{N}_1^n\cos(n\chi)}{1+\mathcal{N}_1^{2}-2\mathcal{N}_1\cos(\chi)}\right)
\end{eqnarray}
Since application of $\hat{T}$ (which is truncated) within this approximation changes the norm of the state, we normalize the state at any step as:
\begin{eqnarray}\label{eq:approx_psin}
|\Psi_n\ra=\frac{\mathcal{N}_1^n}{\sqrt{\mathcal{N}_1^{2n} + 2L\delta_x^2|F_n|^2}}|0\ra^{\otimes L}\nonumber\\
+\frac{F_n}{\sqrt{\mathcal{N}_1^{2n} + 2L\delta_x^2|F_n|^2}}|\delta\ra.
\end{eqnarray}
We note that this state has net zero total magnetization, since:
\begin{eqnarray}
\la\Psi_n|\sigma_j^z|\Psi_n\ra=0\hspace{10pt}\forall j.
\end{eqnarray}

\begin{figure}[!ht]
    \centering
    \includegraphics[scale=0.7]{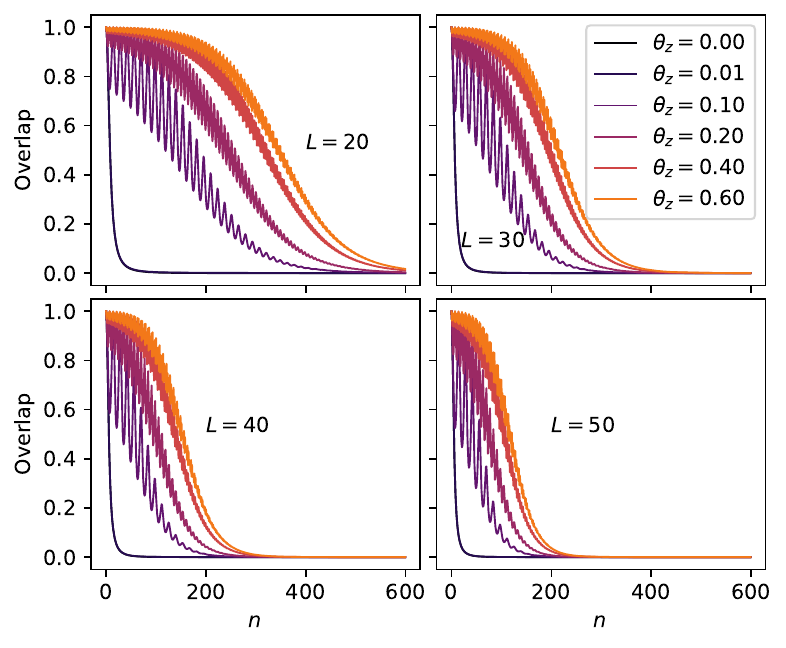}
    \caption{Approximate overlap calculations corresponding to Eq.~\eqref{eq:fide_pert}, for different system sizes $L$ and $\theta_z$ with $\theta_x=0.04$ and $\epsilon=0$. Increasing $\theta_z$ leads to longer-lived return probability as reported in the main text.}    
    \label{fig:fig_s5}
\end{figure}

Eq.~\eqref{eq:approx_psin} provides an approximate expression for the overlap with respect to the initial state:
\begin{eqnarray}\label{eq:fide_pert}
    \mathcal{F}_{p}&=&\left|\langle\Psi_0|\Psi_n\rangle\right|^2=\frac{1}{1 + x^2}\leq 1, \nonumber\\
    x &=&\sqrt{2L}\left(\frac{\delta_x |F_n|}{\mathcal{N}_1^n}\right),
\end{eqnarray}
with $\mathcal{N}_1$ and $F_n$ defined in Eqs.~\eqref{eq:N1_norm} and~\eqref{eq:Fn_norm}, respectively. Eq.~\eqref{eq:fide_pert} introduces a large number of errors at each application of $U_F$, since any term of order $\mathcal{O}(\delta_x^2)$ or higher is iteratively discarded. Moreover, we see that for $L\to \infty$, the overlap rapidly decays to 0, meaning the approach is not suitable for describing results in the thermodynamic limit and higher order terms in $\delta_x$ need to be retained in the expansion in order to capture the correct system-size scaling. Yet, we observe that with this simple approach, overlaps get enhanced with increasing $\theta_z$ values. The overlaps obtained from Eq.~\eqref{eq:fide_pert} are represented in Fig.~\ref{fig:fig_s5}.

\begin{figure}[!ht]
    \centering
    \includegraphics[scale=0.7]{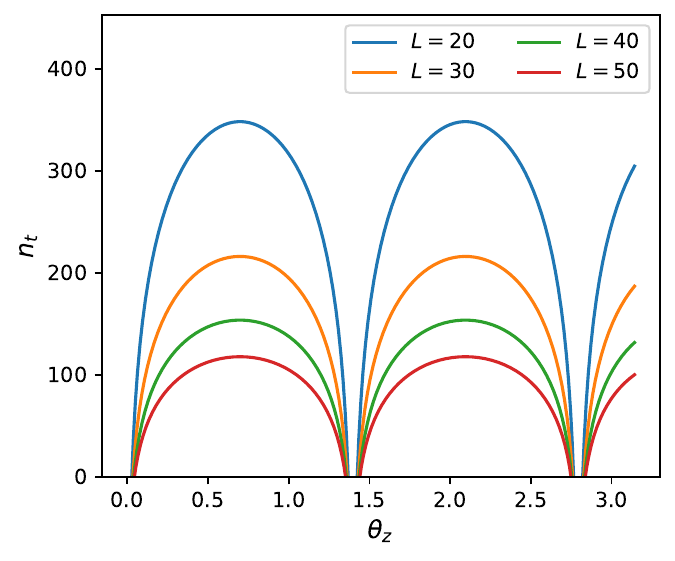}
    \caption{Analytic estimate for the time scale $n_t$ at which the overlap with respect to the initial state in Eq.~\eqref{eq:fide_pert} becomes $\frac{1}{2}$, for different system sizes $L$ and $\theta_z$, with $\theta_x=0.04$ and $\epsilon=0$. Valid solutions are obtained for a wide range of $\theta_z$. Within this approximation, two of the dip regions are in accordance with those observed at points $\theta_z=\frac{4\pi}{9},\frac{8\pi}{9}$ in the phase diagram in the main text.}    
    \label{fig:fig_s6}
\end{figure}

If we focus on the point $x=1$, we can rewrite the equation as (defining $\mathcal{N}_1=\eta$):
\begin{eqnarray}
    \left(\frac{1-\eta^2}{\eta^{2n}}\right)\left(\frac{1+\eta^{2n} -2\eta^n\cos(n\chi)}{1+\eta^2-2\eta\cos(\chi)}\right)=1.
\end{eqnarray}
We can estimate the time scales (time step $n_t$) at which this happens, the associated $n_t$ will be approximately given by the integer part of:
\begin{eqnarray}
n_t&\sim&\frac{\log(1-\eta^2)-\log(1+\eta^2-2\eta\cos(\chi))}{2\log\eta},\nonumber\\
\eta&=&\frac{1}{\sqrt{1 + 2L\delta_x^2}}.
\end{eqnarray}
The estimated values of $n_t$ are represented in Fig.~\ref{fig:fig_s6} for different system sizes. As it has been already discussed, the results do not apply in the thermodynamic limit. However, the overall overlap enhancement with finite $\theta_z$ is clearly visible in this regime. Moreover, the analytic estimates for the time scale $n_t$ are in accordance with the observed overlap phase diagram in the main text, where some of the dips at finite $\theta_z$ correspond to values:
\begin{eqnarray}\label{eq:thetaz_dip1}
    \theta_z=\frac{2\pi m}{9},\hspace{10pt}m=2,4.
\end{eqnarray}
However, this simple derivation does not explain the dip taking place at $\theta_z\sim 2\pi/3$ in the phase diagram, which might need higher order corrections in the overlap calculation, i.e. it might correspond to a non-perturbative point.

\section{Derivation of the effective Floquet Hamiltonian}\label{app:effh}

\begin{figure}[!ht]
    \centering
    \includegraphics[width=0.68\textwidth]{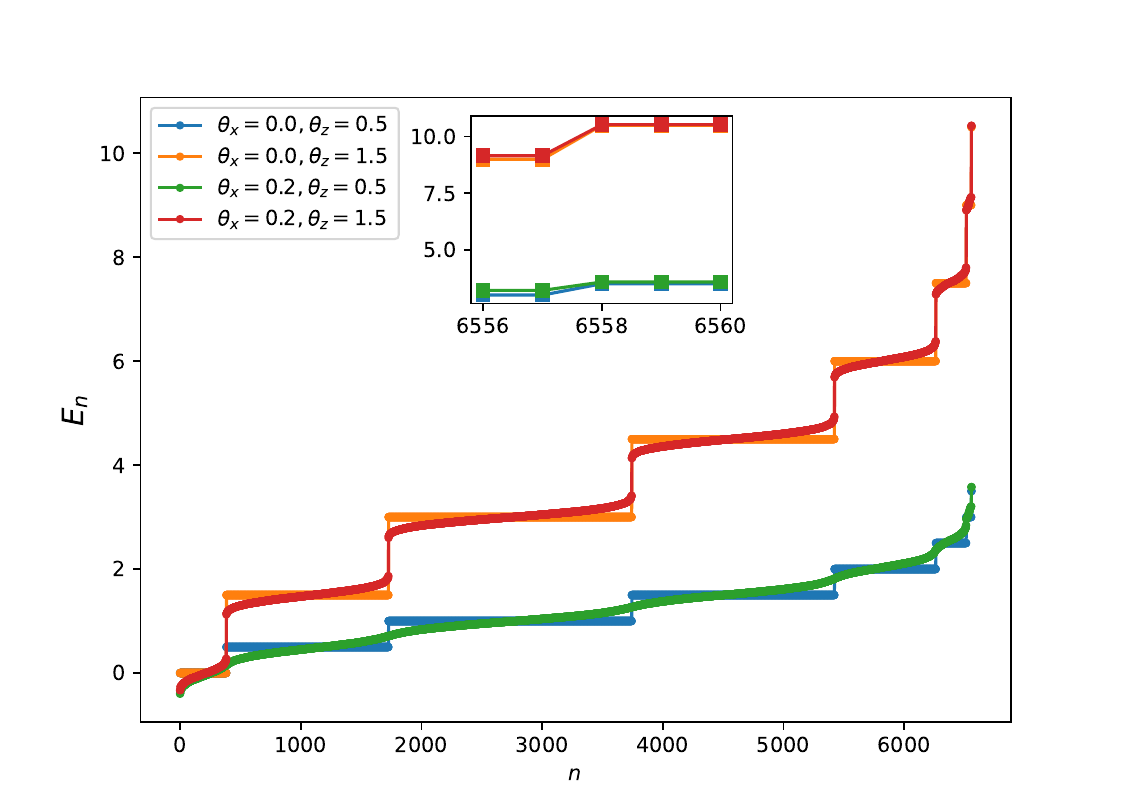}
    \caption{Spectrum of the effective Floquet Hamiltonian for $L=8$ chain with OBC and different values of $\theta_x,\theta_z$. The inset is a zoom in the region with largest eigenvalues, wich corresponds to the uniform product states. Even for finite $\theta_x$, these states remain gapped from the rest of the spectrum. Note that for large $\theta_z=1.5$, more sectors of the Hilbert space remain protected even for finite $\theta_x$. }  
    \label{fig:fig_s7}
\end{figure}

Consider a time-periodic Hamiltonian with period $T$:
\begin{eqnarray}
H(t)=H_0(t)+V(t).
\end{eqnarray}
We consider $H_0(t)$ to be of the form:
\begin{eqnarray}
H_0(t)=\sum_{k=-\infty}^{+\infty}\delta(t-kT)\sum_{l,j} c_{l,j}\lambda_{l,j},
\end{eqnarray}
where $\lambda_{l,j}$ is the $l$ Gell-Mann matrix acting on site $j$. The coefficients $c_{l,j}$ are chosen such that:
\begin{eqnarray}
U_0(kT,0)=\mathcal{T}e^{-i\int_0^{kT} d\tau H_0(\tau)}=
\left(P^{\mathbb{Z}_3}_{\epsilon=0}\right)^k\coloneqq P^k.
\end{eqnarray}

For the interacting part $V(t)$, we write ($k\geq 1,0<T_1<T$):
\begin{eqnarray}
V(t)=\begin{cases}
J_x\sum_j S_j^x S_{j+1}^{x} \hspace{10pt} (k-1)T<t<kT_1\\
J_z\sum_j S_j^z S_{j+1}^{z} \hspace{10pt} kT_1< t< kT,\\
\end{cases}
\end{eqnarray}
For sufficiently fast applications of the drive ($T\lambda\ll 1$, where $\lambda$ is a characteristic energy scale of $V(t)$)  we can estimate the form of the effective Floquet Hamiltonian~\cite{Else2017prx} to be given by:
\begin{eqnarray}
\bar{V}=\frac{1}{NT}\int_0^{NT}dt V_{\text{int}}(t), \hspace{10pt}V_{\text{int}}(t)=U_0^\dagger(t,0)V(t)U_0(t,0).
\end{eqnarray}
Thus, the dynamics of the system due to the Floquet operator $U_F$ can be approximately described by the time-independent unitary given by:
\begin{eqnarray}
U_F\approx P e^{-iT\bar{V}}.
\end{eqnarray}
Then in the interaction picture, for the delta-like pulses considered above, we can split these contributions as:
\begin{eqnarray}
\bar{V}=\frac{1}{3T}\int_0^T dt V(t) + \frac{1}{3T}\int_T^{2T} dt P^\dagger V(t) P + \frac{1}{3T}\int_{2T}^{3T} dt (P^\dagger)^2  V(t) P^2.
\end{eqnarray}
We write now the three different contributions using $H_{\alpha=x,z}=\sum_j S_j^\alpha S_{j+1}^\alpha$:
\begin{eqnarray}
\bar{V}_1 &=& \frac{T_1 J_x}{3T}H_x + \frac{(T-T_1) J_z}{3T}H_z\nonumber\\
 \bar{V}_2 &=& \frac{T_1 J_x}{3T}P^\dagger H_x P + \frac{(T-T_1) J_z}{3T}P^\dagger H_z P\nonumber\\
\bar{V}_3 &=& \frac{T_1 J_x}{3T}(P^\dagger)^2 H_x P^2 + \frac{(T-T_1) J_z}{3T}(P^\dagger)^2 H_z P^2.
\end{eqnarray}
Note now that all terms contribute with equal weight. Consider now the contributions from the terms proportional to $J_z$. We obtain:
\begin{eqnarray}
\bar{V}_z = \frac{(T-T_1)J_z}{T}\sum_j P_{j,j+1} -  \frac{(T-T_1)J_z}{3T}\sum_j I,
\end{eqnarray}
where we have defined:
\begin{eqnarray}
P_{j,j+1}=\sum_{\sigma=+,0,-}|\sigma\sigma\rangle\langle \sigma\sigma|.
\end{eqnarray}
The second term is a constant that can be ignored for the dynamics. The first term  shows that uniform states $|\sigma\rangle^{\otimes L}$ are energy eigenstates for $J_x = 0$ and are in fact three-fold degenerate, while any other product states in the $z$-basis will correspond to a higher degenerate sector. In essence, this shows that $\bar{V}_z$ is gapped, with the value of the gap given by (assuming periodic boundary conditions on a chain of $L$ spins):
\begin{eqnarray}
\Delta _{\bar{V}_z}=\frac{2(T-T_1)J_z}{T}=\frac{2\theta_z}{T},
\end{eqnarray}
where in the last step we have defined $\theta_z=(T-T_1)J_z$. In addition, in the absence of perturbations from the $H_x$ terms, the uniform states are symmetry protected by the $\mathbb{Z}_3$ symmetry.

In order to obtain the contributions from the terms proportional to $J_x$ in the effective Hamiltonian, we employ the following relations of the Gell-Mann matrices and the spin-1 operators:
\begin{eqnarray}
P^\dagger S^x P &=&\frac{\lambda_1-\lambda_5}{\sqrt{2}} =\frac{1}{2}\left[S^x + \{S^x,S^z\} \right]- \frac{1}{\sqrt{2}}\{S^x,S^y\}\nonumber\\
(P^\dagger)^2 S^x P^2 &=&\frac{\lambda_6-\lambda_5}{\sqrt{2}} =\frac{1}{2}\left[S^x - \{S^x,S^z\} \right]- \frac{1}{\sqrt{2}}\{S^x,S^y\},
\end{eqnarray}
where we used that:
\begin{eqnarray}
\lambda_1 =\left(\begin{matrix}
0&1&0\\
1&0&0\\
0&0&0
\end{matrix}\right),\hspace{5pt}\lambda_6 =\left(\begin{matrix}
0&0&0\\
0&0&1\\
0&1&0
\end{matrix}\right),\hspace{5pt}
\lambda_5 &=&\left(\begin{matrix}
0&0&-i\\
0&0&0\\
i&0&0
\end{matrix}\right).
\end{eqnarray}
In total, the contributions of the effective Hamiltonian proportional to $J_x$ read:
\begin{align}
\bar{V}_x &= \bar{V}_{1}+ \bar{V}_{2}+ \bar{V}_{3},\nonumber\\
\bar{V}_1 &= \frac{\theta_x}{3T} \left(\frac{3}{2}\sum_j S_j^x S_{j+1}^x + \sum_j \{S^x_j, S^y_j\}\otimes \{S^x_{j+1}, S^y_{j+1}\}\right)\nonumber\\
\bar{V}_2 &=-\frac{\theta_x}{3\sqrt{2}T}\left(\sum_j \{S^x_j, S^y_j\}\otimes S_{j+1}^x
+\sum_j S_j^x\otimes \{S^x_{j+1}, S^y_{j+1}\}\right)\nonumber\\
\bar{V}_3&= \frac{\theta_x}{6T} \left(\sum_j\{S^x_j, S^z_{j}\}\otimes \{S^x_{j+1}, S^z_{j+1}\}\right),
\end{align}
where we identified $\theta_z=(T-T_1)J_z$ and $\theta_x=J_x T_1$. The spectrum of the effective Hamiltonian is represented in Fig.~\ref{fig:fig_s7} for some values of $\theta_x,\theta_z$. We observe that large $\theta_z$ values protect several Hilbert space sectors in that they remain gapped. The uniform product states are shown in the inset of Fig.~\ref{fig:fig_s7}.

\section{Quantum Fisher Information}\label{subsec:qfi_details}

We define the total QFI to have contributions from each of the total angular momentum components:
\begin{eqnarray}\label{eq:qfi_def}
F_Q &=&\sum_{l=x,y,z}F_Q^l,\hspace{10pt} F_Q^l = 4(\Delta \hat{J}_l)^2,\nonumber\\
\hspace{10pt}\hat{J}_l &=&\sum_{j=1}^L S_j^l,
\end{eqnarray}
with the individual spin components $S_j^{l=x,y,z}$ at lattice site $j$ defined in Eq.~\eqref{eq:spin_defs}.

\subsection{Entanglement bounds in the QFI for $S=1$ chains: Separable states criteria}
We consider the QFI for all separable states of spin-1 of the following operators:
\begin{eqnarray}
\hat{J}^l=\sum_{n=1}^L S^l, \hspace{10pt}l=x,y,z.
\end{eqnarray}
According to the theory of detection of entanglement~\cite{Toth2012}, the variance of the collective operator $\hat{J}^l$ is the sum of variances of each spin (in the case of product states):
\begin{eqnarray}
(\Delta \hat{J}^l)^2 = \sum_{n=1}^L (\Delta S_n^l)^2.
\end{eqnarray}
Thus, we can calculate the total variance as:
\begin{eqnarray}
(\Delta \hat{J})^2=\sum_{l=x,y,z}(\Delta \hat{J}^l)^2 = \sum_{l=x,y,z}\sum_{n=1}^L (\Delta S_n^l)^2.
\end{eqnarray}
The variance is to be calculated respect to states in the $z$-basis projection, i.e:
\begin{eqnarray}
(\Delta S^l_n)^2 &=& \langle s| (S^l_n)^2 |s\rangle - \langle s| S^l_n |s\rangle^2,\nonumber\\
|s\rangle &=&|+\rangle\lor|0\rangle\lor|-\rangle.
\end{eqnarray}
Starting with the $z$ components, we have:
\begin{eqnarray}
\langle s| (S^z_n)^2 |s\rangle=\begin{cases}
1\hspace{10pt}|s\rangle=|+\rangle\lor|-\rangle.\\
0\hspace{10pt}|s\rangle=0.
\end{cases}
\end{eqnarray}
For the other components we obtain:
\begin{eqnarray}
\langle s| (S^x_n)^2 |s\rangle=\langle s| (S^y_n)^2 |s\rangle=\begin{cases}
\frac{1}{2}\hspace{10pt}|s\rangle=|+\rangle\lor|-\rangle.\\
1\hspace{10pt}|s\rangle=0.
\end{cases}
\end{eqnarray}
We also have $\langle s|S^x|s\rangle=\langle s|S^y|s\rangle=0$ and $\text{max}(\langle s|S^z|s\rangle)=1$. The upper-bound of product states for spin-1 is then given by:
\begin{eqnarray}
\text{UB}=\sum_{n=1}^L\left( 1+1+0 -0-0-0 \right)=2L.
\end{eqnarray}
Using Eq.~\eqref{eq:qfi_def}, we find that for product states:
\begin{eqnarray}
F_Q(|\psi\rangle,\hat{J})=4(\Delta \hat{J})^2\to F_Q(|\psi\rangle,\hat{J})\leq 8L.
\end{eqnarray}
We can define the scaled QFI $f_Q$ and write:
\begin{eqnarray}
f_Q(|\psi\rangle,\hat{J})=\frac{1}{8L}F_Q(|\psi\rangle,\hat{J}),\hspace{10pt}f_Q(|\psi\rangle,\hat{J})>1,
\end{eqnarray}
for entanglement witnessing, i.e. if $f_Q(|\psi\rangle,\hat{J})>1$ the state is guaranteed to be entangled. For separable states, this quantity is bounded from below and above by:
\begin{eqnarray}\label{eq:prod_states_bounds}
    f_Q(|\psi\rangle,\hat{J}) &\geq& 0.5,\nonumber\\
    f_Q(|\psi\rangle,\hat{J})&\leq& 1.
\end{eqnarray}

\subsection{Bounds on the QFI observable}\label{subsec:qfi_bounds}
Here we extend the treatment for $S=1$ systems to characterize the different multi-particle entanglement bounds. A $k$-producible pure state is a tensor product of at most $k$-spin entangled states. The maximally entangled states saturate the QFI when the variance of total angular momentum reaches its upper bound:
\begin{eqnarray}
F_Q\leq 4\sum_l \langle \hat{J}_l^2\rangle \leq 4L(L+1).
\end{eqnarray}
The second inequality comes from the theory of angular momentum for a system of $L$ spins with spin value $s$, because:
\begin{eqnarray}
    \langle J_x^2 + J_y^2 + J_z^2 \rangle \leq LS(LS+1).
\end{eqnarray}
Thus, the scaled QFI $f_Q$ gets bounded from above for pure states by:
\begin{eqnarray}
f_Q\leq \frac{S}{2}\left(LS+1\right).
\end{eqnarray}

\subsection{Sections of multipartite entanglement}
We turn now to the set of $k$-producible states, focusing on the spin-1 case $S=1$. We consider the set of pure states that can be written as:
\begin{eqnarray}
|\Psi_{k-\text{producible}}\rangle=|\Psi_1^{(N_1)}\rangle\otimes |\Psi_2^{(N_2)}\rangle\otimes...,
\end{eqnarray}
where the $N_m\leq k$. The case $k=1$ is special because it corresponds to the separable states. One can now write the variances for each of the entangled states:
\begin{eqnarray}
F_Q=4\sum_m (\Delta J)^2_{|\Psi_m^{(N_m)}\rangle}\leq \sum_m 4N_m(N_m+1).
\end{eqnarray}
If $k$ is a divisor of $L$, then the largest variance corresponds to having $N_m=k$ for any $m$. Thus, we can write:
\begin{eqnarray}
F_Q\leq 4 \left[\frac{L}{k}\right]k(k+1)=4 nk(k+1),
\end{eqnarray}
where $[]$ denotes the integer part. If $k$ is not divisor of $L$, then at least one of the states will have fewer than $k$ spins, in which case the maximum of the variance is obtained if all but one state have $k$-spins. Using the form above, we can divide this as:
\begin{eqnarray}
F_Q(k)\leq 4nk(k+1) + 4(L-nk)(L-nk+1),
\end{eqnarray}
which leads to the following bounds for $k$-producible $S=1$ states:
\begin{eqnarray}
f_Q(k)\leq \frac{nk(k+1)}{2L}+\frac{(L-nk)(L-nk+1)}{2L}.
\end{eqnarray}

\section{Experimental Setup}\label{sec:exp_setup}

The experiments are performed using four ${}^{40}\textrm{Ca}^+$ ions confined in a linear Paul ion trap. The computational levels forming the qutrit are encoded in the Zeeman sub-levels of the ground ${}^2S_{1/2}$ and metastable ${}^2D_{5/2}$ states, identifying $|+\rangle={}^2D_{5/2}|m_J = -3/2\rangle$, $|0\rangle={}^2S_{1/2}|m_J = -1/2\rangle$ and $|-\rangle= {}^2D_{5/2}|m_J = -1/2\rangle$. State preparation via optical pumping, Doppler cooling and state-selective fluorescence detection are performed using a laser at 397~nm, with additional lasers at 854~nm and 866~nm to pump the ion out of the metastable ${}^2D_{5/2}$ and ${}^2D_{3/2}$ levels respectively. 

The single- and two-qubit gates are performed on the optical quadrupole ${}^2S_{1/2} \leftrightarrow {}^2D_{5/2}$ transition, using a 729~nm stabilized semiconductor laser. All gates are realized between two sub-levels of the qutrit; single-qubit $\sigma_x$ and $\sigma_y$ gates are driven resonantly, $\sigma_z$ phase shifts are implemented noise-free in software, and two-qubit entangling gates are achieved using the M{\o}lmer-S{\o}rensen interaction. Full details can be found in Ref.~\cite{Ringbauer2022}.

\section{Details on the implementation on the ion trap quantum computer}\label{sec:model_quditproc}
Here we provide details on how to implement the Floquet unitary evolution operator employing the native gate set of the ion-trap universal qudit quantum processor in~\cite{Ringbauer2022}. From now on, all spins-1 of the chain are regarded as \emph{qutrits}, i.e. three level quantum systems. Although we have included explicit expressions for the gates and operations, we refer to the original work Ref.~\cite{Ringbauer2022} and its Supplemental material for complementary purposes. 

The native single qutrit gates can be expressed in terms of the Gell-Mann matrices as:
\begin{eqnarray}
R_1(\theta,\phi) &=& e^{-\frac{\mathrm{i}\theta}{2}\left(\cos(\phi)\lambda_1 + \sin(\phi)\lambda_2\right)}=R_{+0}(\theta,\phi),\nonumber\\
R_2(\theta,\phi) &=& e^{-\frac{\mathrm{i}\theta}{2}\left(\cos(\phi)\lambda_4 + \sin(\phi)\lambda_5\right)}=R_{+-}(\theta,\phi),\nonumber\\
R_3(\theta,\phi) &=& e^{-\frac{\mathrm{i}\theta}{2}\left(\cos(\phi)\lambda_6 + \sin(\phi)\lambda_7\right)}=R_{0-}(\theta,\phi).
\end{eqnarray}
   
The native qutrit-qutrit gate, which defines an entangling operation, is given in terms of the Gell-Mann matrices $\lambda_1,\lambda_2$. Defining:
\begin{eqnarray}
\Lambda_\phi=\cos(\phi)\lambda_1 + \sin(\phi)\lambda_2,
\end{eqnarray}
we have the qutrit-qutrit gate as:
\begin{eqnarray}
\text{MS}(\theta,\phi)=e^{\mathrm{i}\frac{\theta}{2}}e^{-\ii\frac{\theta}{4}(\Lambda_\phi\otimes 1 + 1\otimes\Lambda_\phi)^2 }.
\end{eqnarray}   
We note that for $\phi=0$, we can write the two-site unitary as:
\begin{eqnarray}
\text{MS}(\theta,\phi=0) &=&D -\mathrm{i}\sin(\theta/2)\lambda_1\otimes\lambda_1,
\end{eqnarray}
where $D$ is a diagonal matrix. For the particular case when $\phi=0$, the whole expression factorizes into:
\begin{eqnarray}
\text{MS}(\theta,\phi=0)=e^{\ii\frac{\theta}{2}}e^{-\ii\frac{\theta}{4}(\lambda_1^2\otimes 1)}e^{-\ii\frac{\theta}{4}(1\otimes \lambda_1^2)}e^{-\ii\frac{\theta}{2}(\lambda_1\otimes \lambda_1)}.\nonumber
\end{eqnarray}
The first goal of the approach is to get rid of the unwanted, local rotation terms (see Figs.~1{\bf{c}}, {\bf{d}} in the main text):
\begin{eqnarray}
e^{-\ii\frac{\theta}{4}(\lambda_1^2\otimes 1)}e^{-\ii\frac{\theta}{4}(1\otimes \lambda_1^2)}.
\end{eqnarray}
To get rid of such terms, individual qutrits are encoded into a higher, 4-dimensional space.

\subsection{Qutrit encoding}
Following the Supplemental Information in~\cite{Ringbauer2022}, we show how to encode a qutrit employing an extended local Hilbert space with an auxiliary level that collects unwanted local phases. In order to get rid of the unwanted local phases, we aim at isolating single phase factors into an auxiliary subspace of a single qutrit. To this end, we encode a qutrit into a 4-level system, which can be done for instance using levels $|0\rangle,|1\rangle,|2\rangle,|3\rangle$ in the ion-trap, assigning each level to $|a\rangle,|+\rangle,|0\rangle,|-\rangle$ (see Fig.~1{\bf{b}} in the main text). The local Hilbert space is now of dimension $4\times 4$. For example, now we represent the $\lambda_1$ matrix as (order of the basis is $|a\rangle,|+\rangle,|0\rangle,|-\rangle$ from left to right):
\begin{eqnarray}
\lambda_{1}=\left(\begin{matrix}
0&0&0&0\\
0&0&1&0\\
0&1&0&0\\
0&0&0&0
\end{matrix}\right).
\end{eqnarray}

\subsection{Local rotations}
We employ the following decomposition for the $Z$ gate between levels $01(+0)$ of the qutrit (the rotation matrices are defined above):
\begin{eqnarray}
Z_{+0}(\theta)&=&R_1(\pi/2,-\pi/2)R_1(\theta,0)R_1(\pi/2,\pi/2)\nonumber\\
&=&\left(\begin{matrix}
e^{+\ii\frac{\theta}{2}}&0&0\\
0&e^{-\ii\frac{\theta}{2}}&0\\
0&0&1
\end{matrix}\right).
\end{eqnarray}
In the encoded qutrit space, this extends to the following $4\times 4$ matrix.
\begin{eqnarray}
Z_{+0}(\theta)&=&R_1(\pi/2,-\pi/2)R_1(\theta,0)R_1(\pi/2,\pi/2)\nonumber\\
&=&\left(\begin{matrix}
1&0&0&0\\
0&e^{+\ii\frac{\theta}{2}}&0&0\\
0&0&e^{-\ii\frac{\theta}{2}}&0\\
0&0&0&1
\end{matrix}\right).
\end{eqnarray}
From now on, all single-qutrit gates are understood to be defined in this extended $4\times 4$ space.

Thus, we can partially get rid of local terms as follows:
\begin{eqnarray}\label{eq:local_rot}
&\to&\left[\underbrace{Z_{a+}(-\theta)Z_{+0}(-\theta/2)}_{\text{qutrit} 1}\otimes \underbrace{Z_{a+}(-\theta)Z_{+0}(-\theta/2)}_{\text{qutrit}2}\right]\nonumber\\
&\times &\left[e^{\ii\frac{\theta}{2}}e^{-\ii\frac{\theta}{4}\left[(1\otimes \lambda_1^2)+(\lambda_1^2\otimes 1)\right]}e^{-\ii\frac{\theta}{2}(\lambda_1\otimes \lambda_1)}\right]\nonumber\\
&=&e^{\ii\frac{\theta}{2}}A_1 A_2 e^{-\ii\frac{\theta}{2}(\lambda_1\otimes \lambda_1)},
\end{eqnarray}
where the local rotations get rid of the unwanted phase factors entering the $|+\rangle\langle +|$ components, leaving $A_1,A_2$ as local operators that include a phase factor only in the auxiliary level of qutrit 1 and 2, respectively. Explicitly, $Z_{ij}(\theta)$ represents a rotation between levels $i,j$. The $A_1,A_2$ matrices have the form:
\begin{eqnarray}
A_1=\left(\begin{matrix}
e^{-\ii\frac{\theta}{2}}&0&0&0\\
0&1&0&0\\
0&0&1&0\\
0&0&0&1
\end{matrix}\right).
\end{eqnarray}
Since any other operations carried out in the state will depend only on projectors $|s\rangle\langle s'|,\{s,s'\}\in\{+,0,-\}$, i.e. of the physical qutrit subspace, the matrices $A_1,A_2$ commute with any other operation in that subspace, and do not influence the dynamics of the system.

\subsection{Engineering spin-spin interactions}
We come now to the task of engineering spin interactions. To this end, we look for a unitary transformation between the $\lambda_1$ that leads us directly to $S^z$. Some useful relations are:
\begin{eqnarray}
R_2(\pi/2,\pi/2)\lambda_1 R_2^\dagger(\pi/2,\pi/2)&=&S_{x},\nonumber\\
R_1(-\pi/2,+\pi/2)\lambda_1 R_1(-\pi/2,+3\pi/2)&=&\lambda_3.
\end{eqnarray}
We introduce the following notation for operators acting on the composite Hilbert space $\mathcal{H}=\mathcal{H}_1\otimes \mathcal{H}_2$. We define an operator acting on both subspaces as:
\begin{eqnarray}
O^{12}\equiv O_1\otimes O_2.
\end{eqnarray}
In all cases, the transformations assume that the gate $e^{-\ii\frac{\theta}{2}(\lambda_1\otimes \lambda_1)}$ has been realized using the local rotations described in Eq.~\eqref{eq:local_rot}.

\subsubsection{Local $zz$ interactions}

The $\lambda_1,\lambda_3$ Gell-Mann matrices are related by a unitary transformation:
\begin{eqnarray}
&\to&\left[R_1(-\pi/2,+\pi/2)\otimes R_1(-\pi/2,+\pi/2)\right]e^{-\ii\frac{\theta_z}{2}\lambda_1\otimes\lambda_1} \nonumber\\
&\times&\left[R_1(-\pi/2,+3\pi/2)\otimes R_1(-\pi/2,+3\pi/2)\right]\nonumber\\
&=&e^{-\ii\frac{\theta_z}{2}\lambda_3\otimes\lambda_3}.  
\end{eqnarray}
The $\lambda_3$ matrices can be transformed into the $S^z$ operators as follows (with $\phi=+\pi/2$):
\begin{eqnarray}
\left(
\begin{matrix}
1&0&0\\
0&0&e^{-\ii\phi}\\
0&e^{+\ii\phi}&0
\end{matrix}\right)\left(
\begin{matrix}
1&0&0\\
0&-1&0\\
0&0&0
\end{matrix}\right)\left(
\begin{matrix}
1&0&0\\
0&0&e^{+\ii\phi}\\
0&e^{-\ii\phi}&0
\end{matrix}\right)=S^z.\nonumber
\end{eqnarray}

The rotation matrices correspond to $R_{0-}(+\pi,\pm \frac{\pi}{2})$, which are then applied to the left and right of the previous transformation. The full transformation is then:
\begin{eqnarray}
&\to&R_{0-}^{12}(\pi,\pi/2)R_{+0}^{12}(-\pi/2,+\pi/2)\left[e^{-\ii\frac{\theta_z}{2}\lambda_1\otimes\lambda_1}\right]\nonumber\\
&\times & R_{+0}^{12}(-\pi/2,+3\pi/2)R_{0-}^{12}(\pi,-\pi/2)\nonumber\\
&=&e^{-\ii\frac{\theta_z}{2}S^z\otimes S^z}.
\end{eqnarray}
This way, the local spin-1 interactions have been engineered. 

\subsubsection{Local $xx$ interactions}

The exchange interaction term of the form $S^x_{j+1} S^x_{j}$ can be easily engineered from the entangling gate as:
\begin{eqnarray}
R_{+-}^{12}(\pi/2,\pi/2)e^{-\ii\frac{\theta_x}{2}\lambda_1\otimes\lambda_1}\left[R_{+-}^{12}(\pi/2,\pi/2)\right]^\dagger\nonumber\\
=e^{-\ii\frac{\theta_x}{2}S^x\otimes S^x}.
\end{eqnarray}
\subsubsection{Local qutrit rotations}
The set of local rotations is easily defined in terms of the native single-qutrit gates:
\begin{eqnarray}\label{eq:Pz3}
P_\epsilon^{\mathbb{Z}_3} =\prod_j e^{-\mathrm{i}\frac{\pi-\epsilon}{2}\lambda_{6,j}}\prod_j e^{-\mathrm{i}\frac{\pi-\epsilon}{2}\lambda_{1,j}}.
\end{eqnarray}

\subsection{Floquet unitary under Trotterization}
To make further notation more clear, we will call the set of entangling unitaries on bond $j,j+1$ as:
\begin{eqnarray}
G_{j,zz}=e^{-\ii\frac{\theta_z}{2}S_j^z\otimes S_{j+1}^z}, \hspace{10pt}G_{j,xx}=e^{-\ii\frac{\theta_x}{2}S_j^x\otimes S_{j+1}^x}.
\end{eqnarray}

We introduce now the Floquet unitary of the model in terms of the gates above:
\begin{eqnarray}
U_F=P_\epsilon^{\mathbb{Z}_3}\left(\Pi_{j} G_{j,zz}(\theta_z)\right)\left(\Pi_{j}G_{j,xx}(\theta_x)\right).
\end{eqnarray}
Note that the gates satisfy:
\begin{align}
\left[G_{j,zz}(\theta_z),G_{k,zz}(\theta_z)\right]&=0\quad \forall j,k,\nonumber\\
\left[G_{j,xx}(\theta_x),G_{k,xx}(\theta_x)\right]&=0\quad \forall j,k,,
\end{align}
but:
\begin{eqnarray}
\left[G_{j,zz}(\theta_z),G_{k,xx}(\theta_x)\right]\neq \quad \forall |j-k|\leq 1.
\end{eqnarray}

\end{document}